\documentclass[11pt,epsf]{article}
\usepackage{amsmath,amssymb}
\usepackage{graphicx}
\usepackage{cite,fancyhdr}
\usepackage{braket}
\usepackage{url}
\usepackage{cancel}
\usepackage{color}
\usepackage[colorlinks=true, linkcolor=black, citecolor=black,
urlcolor=black]{hyperref}
\usepackage[left=2.5cm,right=2.5cm,top=2.5cm,bottom=2.5cm,a4paper]{geometry}

\def\l{\left}
\def\r{\right}
\def\nn{\nonumber}

\allowdisplaybreaks[4]
\begin{document}
\setcounter{footnote}{0}
\setcounter{figure}{0}
\setcounter{table}{0}
\begin{titlepage}
\begin{flushright}
{\tt 
OU-HET-1097\\
}
\end{flushright}
\vskip 1.35cm
\begin{center}
{\Large
{\bf
CP-Violation in a Composite 2-Higgs Doublet Model
}
}
\vskip 1.2cm
Stefania De Curtis$^{a,b}$,
Stefano Moretti$^{c}$,
Ryo Nagai$^{d}$,
and
Kei Yagyu$^d$
\vskip 0.4cm
{\small \it $^a$
Istituto Nazionale di Fisica Nucleare (INFN), Sezione di Firenze, \\
Via G. Sansone 1, 50019 Sesto Fiorentino, Italy}\\[3pt]
{\small \it $^b$
Department of Physics and Astronomy, University of Florence, \\
Via G. Sansone 1, 50019 Sesto Fiorentino, Italy}\\[3pt]
{\small \it $^{c}$
School of Physics and Astronomy, University of Southampton, \\ Highfield, Southampton  SO17 1BJ, United Kingdom}\\ [3pt] 
{\small \it $^d$
Department of Physics, Osaka University, Osaka 560-0043, Japan}  \vskip 0.3cm
(\today)
\vskip 1cm
\begin{abstract}
We study CP-Violation (CPV) in a Composite 2-Higgs Doublet Model (C2HDM) based on the 
global symmetry breaking $SO(6)/[SO(4)\times SO(2)]$, wherein  the strong sector is modeled by a two-site moose structure.
Non-trivial complex phases in the interactions involving fermions in both the elementary and strong sectors can induce CPV in  the Higgs potential as well as the Yukawa coupling parameters. We compute both of the latter and analyse their dependence upon the 
aforementioned complex phases. Finally, we discuss physics observables which are distinctive of this model. 
Even in the simplest case with only one complex phase in the strong sector
we can get significant CPV effects.

\end{abstract}
\end{center}
\end{titlepage}
\section{Introduction}

What is the origin of  Electro-Weak Symmetry Breaking (EWSB) as induced by the Higgs mechanism? This is a paramount question to answer in the context of a fundamental theory in which the scalar potential associated to it is generated dynamically. In fact, in 
the Standard Model (SM), the required potential and Higgs field content triggering  EWSB through a non-zero Vacuum Expectation Value (VEV) of a single scalar field are postulated, rather than derived. A by-product of this is that this simple mechanism cannot then  explain why EWSB occurs at an energy scale much lower than other fundamental scales which exist in Nature, such as the Planck scale ($\sim 10^{18}$\,GeV), where gravity becomes as strong as the other three fundamental forces, or else the energy of 
a Grand Unification Theory (GUT)  ($\sim 10^{16}$\,GeV), where all four forces are expected to start unifying into a single unbroken one. This flaw becomes quantitatively manifest through the fact that  the SM cannot explain why the mass of the  Higgs particle discovered at the Large Hadron Collider (LHC) is so light ($M_H\approx125$ GeV) yet so sensitive to such high scale effects, to the extent that it would grow indefinitely with the energy scale at which it is probed, unless {\sl ad hoc} cancellations are invoked in defining the renormalised value of $M_H$. This drawback of the SM is known as the ``hierarchy problem'', also called ``naturalness problem''. 
The SM implementation of EWSB, thus, is inherently incomplete  and one should conclude that it will need to be extended, in order to describe physics in the Ultra-Violet (UV) domain, i.e., at much higher energy scales than the EW one.

Composite Higgs Models (CHMs) provide an attractive solution to the hierarchy problem \cite{
Kaplan:1983fs, 
Dimopoulos:1981xc, 
Kaplan:1983sm, 
Banks:1984gj, 
Georgi:1984ef, 
Dugan:1984hq, 
Georgi:1984af,
Contino:2003ve}.
Herein, a new QCD-like strong dynamics (henceforth, QCD') is assumed to exist at a multi-TeV scale
and the discovered 125\,GeV Higgs particle is considered to be the pseudo-Nambu-Goldstone Boson (pNGB) emerging from 
 ``chiral symmetry breaking'' in the QCD' sector. 
In a similar fashion to what happens in QCD,
the dimensional transmutation mechanism in the QCD' gauge dynamics can naturally 
realise QCD' confinement at a much lower energy scale than the UV cutoff scale. 
Furthermore, an (approximate) shift symmetry protects the Higgs mass from  
large UV radiative corrections in analogy to the pion mass in QCD.
Hence, it is not surprising that CHMs have gathered a lot of attention as a class of leading candidates for 
physics Beyond the SM (BSM).

The first step in building a CHM is to specify the pattern of  ``chiral symmetry breaking'' in the QCD' sector. 
Once the breaking pattern, $G\to H$, is determined, one can formulate the low-energy effective theory by using a non-linear sigma model on the coset $G/H$ \cite{Coleman:1969sm, Callan:1969sn}.
The Higgs potential is then radiatively induced by the explicit breaking of the $G$ symmetry,
which can be realised by gauging (a subgroup of) $G$ and/or introducing linear mixing 
parameters between the QCD' and EW sector. 
The minimal framework to achieve successful EWSB is the model based on  $G/H=SO(5)/SO(4)$,
where 4 pNGBs associated with the symmetry breaking behave like the SM Higgs doublet field \cite{Agashe:2004rs}.
This minimal setup and its phenomenological consequences have been thoroughly 
explored in the literature, see e.g. Refs. \cite{Contino:2010rs, Panico:2015jxa} for reviews.

It is also worth considering non-minimal CHMs in which the emerging Higgs sector is extended beyond the minimal one described above.
An attractive non-minimal CHM is the C2HDM of Refs.~\cite{
Mrazek:2011iu, 
Bertuzzo:2012ya,
DeCurtis:2018iqd, 
DeCurtis:2018zvh
}. 
In this C2HDM, 8 pNGBs emerge from the global symmetry breaking in the QCD' sector 
and  behave like two isospin doublet (pseudo)scalar fields at the low-energy scale. 
The C2HDM has the potential to solve not only the hierarchy problem\footnote{Possibly, more elegantly that its counterpart in Supersymmetry, in fact, the so-called Minimal Supersymmetric Standard Model (MSSM) \cite{DeCurtis:2018iqd}.}
but also unresolved issues left in the minimal CHM. 
For example, it is known that an Elementary 2HDM (E2HDM) provides a successful framework 
for EW baryogenesis, which cannot be realised in the minimal Higgs sector \cite{
Nelson:1991ab, 
Turok:1991uc, 
Funakubo:1993jg, 
Davies:1994id, 
Cline:1995dg, 
Kanemura:2004ch, 
Kanemura:2005cj, 
Fromme:2006cm, 
Tulin:2011wi,
Cline:2011mm,
Liu:2011jh,
Chiang:2016vgf,
Guo:2016ixx,
Fuyuto:2017ewj,
Modak:2018csw,
Fuyuto:2019svr}. 
Moreover, 
the E2HDM framework can be also considered as the low-energy effective theory for a complete model explaining the origin of the tiny but non-zero neutrino masses \cite{Zee:1980ai, Aoki:2008av, Aoki:2009vf, Kanemura:2013qva}.
Such attractive features of the E2HDM motivated some of us to consider a C2HDM scenario, which would then inherit these.

Specifically, in Refs.~\cite{DeCurtis:2018iqd, DeCurtis:2018zvh,DeCurtis:2019jwg}, we investigated a C2HDM based on the spontaneous breaking of a $G/H=$ $SO(6)/[SO(4)\times SO(2)]$ global symmetry. 
This may be the most economical setup for generating a C2HDM
since other realisations need a global symmetry $G$ with a larger dimension than $SO(6)$. Here, 
the low-energy effective theory for the new strong dynamics is explicitly described by a two-site moose model, which is made of two sectors:
an ``elementary sector'', including particles whose quantum numbers under the EW gauge symmetry are the same as those of the SM fermions and gauge bosons, plus a ``composite sector'', having new spin-1/2 and spin-1 resonances introduced as multiplets of the global group. 
The mixing between states in these two sectors realises the  so-called ``partial compositeness'' scenario, wherein the gauge sector
mirrors the analogous of $\gamma$-$\rho$ mixing in QCD while for fermions it is nothing but an assumption   that  implies the existence of heavy fermions with  quantum numbers identical to those of SM quarks and leptons\footnote{Hence, some arbitrariness exists in the choice of the SM fermions subject to such a mixing, which in the C2HDM considered here are identified with the third generation only.} (and potentially of also exotic states). In such a C2HDM,  the shape of the Higgs potential is determined by the partial compositeness parameters and the gauge couplings. This is a remarkable difference between such C2HDM and an E2HDM, wherein the Lagrangian parameters (like in the SM) are put in by hand, hence, they are totally undetermined (see Refs.~\cite{DeCurtis:2016scv,DeCurtis:2016tsm,DeCurtis:2016gly,DeCurtis:2017gzi} for initial attempts to constrain these through both theoretical and experimental requirements). 

In our previous works, we computed the Higgs potential in the C2HDM by assuming CP Conservation (CPC).  In this paper, we extend our analysis by including CP Violation (CPV) effects 
which are induced solely by the partial compositeness mechanism. 
As we will show, the dynamics emerging from partial compositeness is described by linear couplings between the fermion fields living in the elementary and composite sectors and their 
 complex phases can induce non-trivial CPV effects in observables at low-energy scales. Furthermore, we will show that, in such a BSM construct, CPV effects eventually appear in both the Yukawa sector and Higgs potential.
With this in mind, ultimately, we will want to explore  the correlations between CPV as appearing in the  visible sector and as generated in the strong one.

This paper is organised as follows.
In Sect.~\ref{sec:model}, we describe the $SO(6)/[SO(4)\times SO(2)]$ dynamics generating our C2HDM with CPV, specifically reporting several analytic expressions for the Higgs potential and top Yukawas in Sect. \ref{sec:HiggsTop}.
(The details of the calculations involved are reported in the Appendices.) 
In Sect.~\ref{sec:Signatures}, we estimate the typical size of CPV effects in this setup and describe some physics observables which can be studied in order to extract a quantitative measurements of those.
Finally, we conclude in Sect.~\ref{sec:Summary}.

\section{The $SO(6)/[SO(4)\times SO(2)]$ Composite Higgs Model}
\label{sec:model}
We consider the spontaneous symmetry breaking of a global symmetry, $SO(6)\to SO(4)\times SO(2)$, at an energy, $f$, which we call compositness scale. This dynamics produces 8 pNGBs, which can be identified as two $SU(2)$ weak doublet Higgs fields. Concretely, the pNGB matrix $U$ can be parameterised as
\begin{align}
U
=
\exp\l(\frac{i}{f}\Pi\r)
\,,\qquad
\Pi
=
-i\l(
\begin{array}{cc}
0_{4\times 4} & \Phi\\
-\Phi^T & 0_{2\times 2}\\
\end{array}
\r)
\,,
\label{eq:U}
\end{align}
where $\Phi=(\phi^{\hat{a}}_1,\phi^{\hat{a}}_2)$, with the two real 4-vectors $\phi^{\hat{a}}_i$ being rearranged as two $SU(2)$ weak doublet Higgs fields, $\Phi_1$ and $\Phi_2$,
\begin{align}
\Phi_i
=
\frac{1}{\sqrt{2}}
\l(
\begin{array}{ccc}
\phi^{2}_i+i\phi^1_i\\
\phi^{4}_i+i\phi^3_i\\
\end{array}
\r)
\,\qquad
(i=1,2)
\,.
\end{align}

The aforementioned spontaneous symmetry breaking and the strong dynamics emerging in the C2HDM can effectively be described by a two-site moose model \cite{Panico:2011pw,DeCurtis:2011yx,DeCurtis:2018iqd, DeCurtis:2018zvh}, which consists of the two sectors, which  preserve a global $G_i=SO(6)\times U(1)_X~(i=1,2)$ symmetry. 
Hereafter, we call the first and second sectors as the ``elementary'' and ``strong'' sectors, respectively. The Lagrangian of the two-site moose model can then be decomposed into two parts, $\mathcal{L}=\mathcal{L}_{\rm{boson}}+\mathcal{L}_{\rm{fermion}}$. 

The Lagrangian of the bosonic sector ($\mathcal{L}_{\rm{boson}}$) is written as 
\begin{align}
\mathcal{L}_{\rm{boson}}
&=
\frac{f^2_1}{4}\mbox{Tr}[D_\mu U_1D^\mu U_1]
+
\frac{f^2_2}{4}\mbox{Tr}[D_\mu \Sigma_2D^\mu  \Sigma_2]
\nn\\
&
-
\frac{1}{4g^2_\rho}(\rho^A)_{\mu\nu}(\rho^A)^{\mu\nu}
-
\frac{1}{4g^2_{\rho X}}(\rho^X)_{\mu\nu}(\rho^X)^{\mu\nu}
-
\frac{1}{4g^2_{A}}(A^A)_{\mu\nu}(A^A)^{\mu\nu}
-
\frac{1}{4g^2_{X}}X_{\mu\nu}X^{\mu\nu}
\,,
\end{align}  
where $G_1$ and $G_2$ are gauged,  $(A^A_\mu, X_\mu)$ and $(\rho^A_\mu, \rho^X_\mu)$ denote the ($SO(6)$, $U(1)_X$) gauge fields with  
($g_{A}$, and $g_X$) with ($g_\rho$, $g_{\rho X}$) being gauge coupling parameters, respectively. The index $A$ labels the adjoint representation of $G$. 
 The SM  gauge bosons are embedded into the $G_1$ gauge fields, $(A^A_\mu, X_\mu)$. (We regard the other components of the $G_1$ gauge fields as spurions.) Furthermore, $U_1$ and $\Sigma_2$ are the pNGB matrix fields which are made of the chiral fields, $U$ (recall Eq.~(\ref{eq:U})):
\begin{align}
&U=U_1U_2\,,~~~
U_i=\exp\l(i\frac{f}{f^2_i}\Pi\r)
\,,~~~
\Sigma_2=U_2\Sigma_0 U^T_2
\,,
\end{align}
with $f^{-2}=f^{-2}_1+f^{-2}_2$ and $\Sigma_0$ being an $SO(4)\times SO(2)$ invariant vacuum, $\Sigma_0=0_{4\times 4}\otimes\,i\sigma_2$.
We note that $U_1$ plays the role of the link field which connects the two sectors while $\Sigma_2$ is a linear (adjoint) representation of $G_2$. The VEV of $U_1$ triggers the spontaneous symmetry breaking of $G_1\times G_2$ to the diagonal component $G$ while the VEV of $\Sigma_2$ accounts for the breaking to $SO(4)\times SO(2)\times U(1)_X$.
This breaking pattern provides 24 pNGBs, 16 of which are absorbed in the longitudinal components of the gauge fields, while the remaining 8 can be identified with Higgs fields.
The covariant derivatives for the pNGB fields are defined as $D_\mu U_1 = \partial_\mu U_1-iA_\mu U_1+iU_1\rho_\mu$ and $D_\mu \Sigma_2 
= \partial_\mu \Sigma_2-i[\rho_\mu, \Sigma_2]$,  where $A_\mu = A^A_\mu T^A+X_\mu T^X$ and $\rho_\mu = \rho^A_\mu T^A+\rho^X_\mu T^X$, with $T^A$ and $T^X$ being the generators of $SO(6)$ and $U(1)_X$, respectively. We thus note that there is no source of CPV in the bosonic sector Lagrangian.

We next discuss the fermionic sector Lagrangian ($\mathcal{L}_{\rm{fermion}}$). We introduce two fermions, $(q^{\bf{6}}_{L})_t$ and $t^{\bf{6}}_{R}$, in the elementary sector, which are a $SO(6)$ ${\bf{6}}$-plet with $U(1)_X$ charge $X=2/3$. The third generation of the left-handed quark doublets $q_L$ and right-handed top quark $t_R$ are embedded into the $SO(6)$ ${\bf{6}}$-plet fermions as
\begin{align}
&(q^{\bf{6}}_{L})_t
= (\Upsilon^{t}_L)^Tq_L 
\,,~~~
t^{\bf{6}}_{R}
= 
(\Upsilon^{t}_R)^Tt_R \,,
\end{align}
where
\begin{align}
&\Upsilon^{t}_L=
\frac{1}{\sqrt{2}}\l(
\begin{array}{cccccc}
0 & 0 & 1 & i & 0 & 0\\
1 & -i & 0 & 0 & 0 & 0\\
\end{array}
\r)\,,\label{eq:UpsilontL}\\
&\Upsilon^{t}_R=
\l(
\begin{array}{cccccc}
0 & 0 & 0 & 0 & \cos\theta_{t} & i\sin\theta_{t}\\
\end{array}
\r)\,,\label{eq:UpsilontR}
\end{align}
with $\theta_{t}$ being a free parameter valued between $-\pi$ and $\pi$. Hereafter, we consider only the top quark contribution, because the other SM fermions
provide only sub-leading corrections to our later analysis, however, they can be included, if necessary, by simply extending the formalism described above.
We note that $\Upsilon^t_L$ does not bring a new source of CPV, because it corresponds to an Hermite operator that projects out a $SO(6)$ generator to a $SU(2)_L$ one, while $\Upsilon^t_R$ does. As we will see below, a non-zero value of $\sin\theta_t$ can induce CPV.
We also introduce $N$ $SO(6)$ ${\bf{6}}$-plet fermions with $U(1)_X$ charge $X=2/3$, $\Psi^I~(I=1,2,\cdots,N)$, in the strong sector, in order to describe spin-1/2 resonances. 
The $G_1 \times G_2$ invariant Lagrangian of the fermion sector is given as
\begin{align}
\mathcal{L}_{\rm{fermion}}
&=
(\bar{q}^{\bf{6}}_{L})_t
 i\gamma^\mu D_\mu (q^{\bf{6}}_{L})_t
+
\bar{t}^{\bf{6}}_{R} i\gamma^\mu D_\mu t^{\bf{6}}_{R}
+
\bar{\Psi}^I_t i\cancel{\partial} \Psi^I_t
\nn\\
&
-
\bar{\Psi}^I_t
\l[
M^{IJ}_{\Psi} 
+
Y^{IJ}_1\Sigma_2
+
Y^{IJ}_2\Sigma^2_2
\r]P_R\Psi^J_t+h.c.\nn\\
&
+(\bar{q}^{\bf{6}}_{L})_t U_1\Delta^I_L  P_R \Psi^I_t 
+\bar{t}^{\bf{6}}_{R} U_1 \Delta^I_R P_L \Psi^I_t + h.c.
\,,
\label{eq:Lfermi}
\end{align}
where the covariant derivatives of the elementary fermions include the interactions with the elementary gauge bosons while the covariant derivative of the spin 1/2 resonance $\Psi$ provides the couplings to the spin-1 resonances introduced above. We do not write down $(\Sigma_2)^{n}~(n\geq3)$ terms because of $(\Sigma_2)^3=-\Sigma_2$. It should be noted that the parameters in Eq.~(\ref{eq:Lfermi}), $M_{\Psi}, Y_1, Y_2, \Delta_L, \Delta_R$, 
can be taken as complex matrices/vectors and can thus be sources  the CPV in our setup.

The low-energy effective Lagrangian can be obtained by integrating out the spin-1 ($\rho^A_\mu,\rho^X_\mu$) and spin-1/2  ($\Psi$) resonances. The quadratic terms of the SM fermions and gauge bosons in  momentum ($p$) space are written as
\begin{align}
\mathcal{L}_{\rm{eff}}
&=
-\frac{1}{2}\l(\eta^{\mu\nu}-\frac{p^\mu p^\nu}{p^2}\r)\biggl[
p^2\tilde{\Pi}_X(p^2)X_\mu X_\nu
+
p^2\tilde{\Pi}_0(p^2)A^A_\mu A^B_\nu\nn\\
&
+f^2\tilde{\Pi}_1(p^2)A^A_\mu A^B_\nu\mbox{Tr}\l[\Sigma T^AT^B\Sigma\r]+f^2\tilde{\Pi}_2(p^2)A^A_\mu A^B_\nu\mbox{Tr}\l[ T^A\Sigma T^B\Sigma\r]\biggr]
\nn\\
&+
(\bar{q}^{\bf{6}}_L)_t\cancel{p}\l[
\tilde{\Pi}^{q}_0(p^2)
+
\tilde{\Pi}^{q}_1(p^2)\Sigma
+
\tilde{\Pi}^{q}_2(p^2)\Sigma^2
\r]({q}^{\bf{6}}_L)_t\nn\\
&+
\bar{t}^{\bf{6}}_R\cancel{p}\l[
\tilde{\Pi}^{t}_0(p^2)
+
\tilde{\Pi}^{t}_1(p^2)\Sigma
+
\tilde{\Pi}^{t}_2(p^2)\Sigma^2
\r]{t}^{\bf{6}}_R\nn\\
&+
(\bar{q}^{\bf{6}}_L)_t\l[
\tilde{M}^{t}_0(p^2)
+
\tilde{M}^{t}_1(p^2)\Sigma
+
\tilde{M}^{t}_2(p^2)\Sigma^2
\r]{t}^{\bf{6}}_R
+
h.c.
\,,
\label{eq:Leff}
\end{align}
where $\Sigma=U\Sigma_0 U^T = U_1\Sigma_2 U^T_1$. Furthermore, $\tilde{\Pi}$'s and $\tilde{M}$'s are form factors which encode the effect of the strong dynamics. The explicit expressions of the form factors for the $N=2$ case are summarised in  Appendix \ref{eq:formfactors}. We note that the hermeticity of the Lagrangian implies that the form factors $\tilde{\Pi}^q_1$ and $\tilde{\Pi}^t_1$ are purely imaginary, 
while $\tilde{\Pi}^{q_t}_{0,2}$ and $\tilde{\Pi}^{t}_{0,2}$ are real. The form factors $\tilde{M}^t_0$, $\tilde{M}^t_1$ and $\tilde{M}^t_2$
can generally be complex. Therefore, CPV effects should appear in $\tilde{\Pi}^q_1$, $\tilde{\Pi}^t_1$, $\tilde{M}^t_0$, $\tilde{M}^t_1$ and $\tilde{M}^t_2$. 
\section{The Higgs Potential and Top Yukawa Sector}
\label{sec:HiggsTop}
We have introduced the SM gauge bosons and fermions in the elementary sector. 
Since the embedding of the SM fields explicitly breaks the shift symmetry, 
the SM field interactions generate a potential for the pNGB (Higgs) fields. Neglecting $\mathcal{O}(\Phi^6_{i}/f^2)~(i=1,2)$ terms, the Higgs potential can be parameterised in the form of
\begin{align}
V(\Phi_1,\Phi_2)
&=
m^2_1 \Phi^\dag_1\Phi_1
+
m^2_2 \Phi^\dag_2\Phi_2
-
\l[m^2_3\Phi^\dag_1\Phi_2+h.c.\r]\nn\\
&
+
\frac{\lambda_1}{2}(\Phi^\dag_1 \Phi_1)^2
+
\frac{\lambda_2}{2}(\Phi^\dag_2 \Phi_2)^2
+
{\lambda_3}(\Phi^\dag_1 \Phi_1)(\Phi^\dag_2 \Phi_2)
+
{\lambda_4}(\Phi^\dag_1 \Phi_2)(\Phi^\dag_2 \Phi_1)\nn\\
&
+
\l[\frac{\lambda_5}{2}(\Phi^\dag_1 \Phi_2)^2
+
{\lambda_6}(\Phi^\dag_1 \Phi_1)(\Phi^\dag_1 \Phi_2)
+
{\lambda_7}(\Phi^\dag_2 \Phi_2)(\Phi^\dag_1 \Phi_2)
+
h.c.\r]
\,,
\end{align}
where the parameters $m^2_i$ and $\lambda_i$  
are calculated by the momentum integration of the form factors, $\tilde{\Pi}$ and $\tilde{M}$, which are defined in Eq.~(\ref{eq:Leff}). The explicit relations between the Higgs potential parameters and the form factors are summarised in Appendix \ref{app:Higgspotential}. In contrast to the E2HDM, the Higgs potential parameters in the C2HDM are not independently adjustable parameters. Instead, the structure of the Higgs potential is determined by the strong dynamics.
We also note that the C2HDM generally induces non-zero $m^2_3$, $\lambda_6$ and $\lambda_7$ terms, which break (albeit softly) the $\mathbb{Z}_2$ symmetry normally introduced
in the E2HDM to control Flavour Changing Neutral Currents (FCNCs) at tree level~\cite{Glashow:1976nt}.
Thus, the C2HDM generally introduces FCNCs at tree level. 
Such FCNCs can, however, be avoided by imposing Yukawa alignment~\cite{Pich:2009sp}, i.e., two Yukawa matrices are assumed to be proportional with each other. 
Since effects of lighter fermions onto the Higgs potential are negligibly small, we can impose the Yukawa alignment without changing the structure of it.

Let us comment on a technical aspect of the calculation of the Higgs potential. As mentioned in Refs.~\cite{DeCurtis:2018iqd, DeCurtis:2018zvh}, the momentum integration of the form factors generally induces UV divergences. The UV finiteness of the momentum integration requires specific relations among the parameters in the fermion sector, Eq.~(\ref{eq:Lfermi}). We summarise the UV finiteness conditions in  Appendix \ref{app:Higgspotential}.

Depending on the parameters in the strong sector, the Higgs doublet fields can develop non-trivial VEVs. 
Without loss of generality, we can parameterise the VEVs of $\Phi_1$ and $\Phi_2$ as
\begin{align}
\braket{\Phi_1} = \frac{1}{\sqrt{2}}
\begin{pmatrix}
0 \\
v_1
\end{pmatrix},\quad 
\braket{\Phi_2} = \frac{1}{\sqrt{2}}
\begin{pmatrix}
0 \\
v_2\,e^{i\theta_v}
\end{pmatrix}, 
\end{align}
where $v_1$ and $v_2$ are real and positive while $\theta_v\in (-\pi,\pi]$. Here, 
$v_1$, $v_2$ and $\theta_v$ relate to the EW symmetry breaking scale, $v_{\rm{EW}}=(\sqrt{2}G_F)^{-1/2}\simeq 246\,\mbox{GeV}$, with $G_F$ being the Fermi constant, as follows
\begin{align}
v^2_{\rm{EW}}
=
v^2-\frac{v^4}{3f^2}\left(1 -2\sin^2\beta \cos^2\beta \sin^2\theta_{v}\right) + \mathcal{O}(1/f^4). 
\end{align}
where $v=\sqrt{v^2_1+v^2_2}$ and $\tan\beta =
\frac{v_2}{v_1}$.

It is convenient to define the ``Higgs basis'' $(H_1,H_2)$ as
\begin{align}
\begin{pmatrix}
H_1 \\
H_2 
\end{pmatrix}
=
\begin{pmatrix}
\cos\beta & \sin\beta \\
-\sin\beta & \cos\beta
\end{pmatrix}
\begin{pmatrix}
1 & 0 \\
0 & e^{-i\theta_v}
\end{pmatrix}
\begin{pmatrix}
\Phi_1 \\
\Phi_2 
\end{pmatrix}
\,,
\end{align}
so that $H_{1,2}$ satisfy 
\begin{align}
\braket{H_1} = \frac{1}{\sqrt{2}}
\begin{pmatrix}
0 \\
v
\end{pmatrix},\qquad 
\braket{H_2} = 
0\,.
\end{align}
We can easily identify the physical (pseudo)scalar particles in the Higgs doublet fields by taking the Higgs basis. We parameterise the component fields as
\begin{align}
H_1 = \begin{pmatrix}
G^+ \\
\frac{v + \tilde{\phi}^0_1 + iG^0}{\sqrt{2}}
\end{pmatrix},\quad 
H_2 = \begin{pmatrix}
H^+ \\
\frac{\tilde{\phi}^0_2 + i\tilde{\phi}^0_3}{\sqrt{2}}
\end{pmatrix}, \label{eq:components}
\end{align}
where $G^0$ and $G^\pm$ are the Nambu-Goldstone bosons which are absorbed by the longitudinal components of the $Z$ and $W^\pm$ bosons while 
$H^\pm$ and $\tilde{\phi}^0_{1,2,3}$ are the physical charged Higgs boson and neutral Higgs bosons, respectively. 
If the Higgs potential contains a physical CPV phase, these three neutral Higgs bosons $\tilde{\phi}^0_{1,2,3}$ are mixed with each other. 
Their mass eigenstates ($h_1,h_2,h_3$) are defined by introducing a $3\times 3$ orthogonal matrix $R$ as 
\begin{align}
\begin{pmatrix}
\tilde{\phi}^0_{1} \\
\tilde{\phi}^0_{2} \\
\tilde{\phi}^0_{3} 
\end{pmatrix}
=R 
\begin{pmatrix}
H_1 \\
H_2 \\
H_3 
\end{pmatrix}
\label{eq:rmat}
\end{align}
and their masses are defined as 
\begin{align}
R^T{\cal M}_N R = \text{diag}(m_{H_1}^{2},m_{H_2}^{2},m_{H_3}^{2}), \quad (m_{H_1}^{}\leq m_{H_2}^{}\leq m_{H_3}^{}), \label{hessian}
\end{align}
with ${\cal M}_N$ being the Hessian matrix for the  ($\tilde{\phi}^0_{1},\tilde{\phi}^0_{2},\tilde{\phi}^0_{3}$) basis evaluated in vacuum. 
We identify $H_1$ as the discovered Higgs boson at the LHC. 
These masses and the matrix $R$ are calculated via the parameters of the Higgs potential which are in turn determined by the parameters in the strong sector.

The detailed study of the Higgs potential in the C2HDM with CPC was performed in Refs.~\cite{DeCurtis:2018iqd, DeCurtis:2018zvh}. 
Here, let us focus on the CPV terms in the Higgs potential. As we discussed in the previous section, the fermion (top) interactions generally include CPV effects. 
The imaginary part of the Higgs potential parameters are obtained as
\begin{align}
&{\mbox{Im}\l[m^2_3\r]}
=
\frac{6i}{f^2}\int\frac{d^4 p}{(2\pi)^4}
\biggl[
\frac{\mbox{Im}\l[M_2M^*_1\r]}{p^2}
-\frac{\sin2\theta_t}{2}\l(\frac{|M_1|^2+|M_2|^2}{p^2}\r)
+i\Pi^q_1
\biggr]
\,,
\label{eq:Imm3sqwthetat}
\\
&\mbox{Im}\l[{\lambda}_6\r]
=
\frac{6i}{f^4}\int\frac{d^4 p}{(2\pi)^4}
\biggl[
\frac{(4+3{\cos2\theta_t})}{3} \frac{\mbox{Im}\left[{M_2} {M^*_1}\right]}{p^2}
-\frac{2 \sin2\theta_t}{3}\l(\frac{{|M_1|^2+|M_2|^2}}{p^2} \r)
-{i \Pi^q_1} {\Pi^q_2}
+\frac{{i \Pi^q_1}}{3}
\biggr]
\,,
\\
&\mbox{Im}\l[{\lambda}_7\r]
=
\frac{6i}{f^4}\int\frac{d^4 p}{(2\pi)^4}
\biggl[
\frac{(4-3{\cos2\theta_t}) }{3}\frac{\mbox{Im}\left[{M_2} {M^*_1}\right]}{p^2}
-\frac{2  \sin2\theta_t}{3 }\l(\frac{|M_1|^2+|M_2|^2}{p^2}\r)
-i\Pi^q_1 {\Pi^q_2}
+\frac{{i \Pi^q_1}}{3}
\biggr]
\,,
\label{eq:Imlam7wthetat}
\end{align}
where 
\begin{align}
&\Pi^{q}_{1,2}
=
\frac{\tilde{\Pi}^{q}_{1,2}}{\tilde{\Pi}^{q}_0}
\,,
\qquad
M_{1,2}
=
\frac{\tilde{M}_{1,2}}{\sqrt{\tilde{\Pi}^{q}_0(\tilde{\Pi}^{t}_0-\tilde{\Pi}^{t}_2+i\sin2\theta_t\tilde{\Pi}^t_1)}}
\,.
\label{eq:wotildeformfac}
\end{align}
In our analysis, we neglect terms proportional to $(M_i^*M_j)\Pi_{k}^{q,t}$ and $(M_i^*M_j)(M_k^*M_l)$ $(i,j,k,l = 1,2)$. 
These terms are highly suppressed by the factor of $(\Delta_{L,R}/M_*)^6$ and $(\Delta_{L,R}/M_*)^8$ with $M_*$ being a typical mass scale of spin 1/2 resonances, respectively, 
where the ratio $\Delta_{L,R}/M_{*}$ should be smaller than 1 in order to reproduce the top Yukawa coupling, see Eq.~(\ref{eq:mt2}).
With this analysis, we find that $\mbox{Im}[\lambda_5]$ is negligibly small in the C2HDM with CPV.
It should also be noted that one of the three complex phases in Eq.~(\ref{eq:Imm3sqwthetat})--(\ref{eq:Imlam7wthetat}) can be eliminated by using the vacuum condition. 
The remaining physical complex phases may be defined as \cite{Botella:1994cs}
\begin{align}
J_1=\mbox{Im}\l[\frac{m^2_3}{f^2}\,{\lambda}^*_6\r]\,,\qquad
J_2=\mbox{Im}\l[\frac{m^2_3}{f^2}\,{\lambda}^*_7\r]\,. \label{eq:invariance}
\end{align}
The other possible quantities carring CPV, such as $\mbox{Im}[\lambda_6{\lambda}^*_7]$, can be expressed in terms of  $J_1$ and $J_2$.
We then observe that a non-zero $\sin2\theta_t$, ${\Pi}^q_{1}$ and/or $\mbox{Im}[{M}_2{M}^*_1]$ onsets CPV in the Higgs sector. 

Let us now discuss some phenomenological aspects of  CPV in the Higgs sector of our C2HDM, also in relation to the gauge one. 
There are common and different aspects when comparing it to the E2HDM with CPV.  The common one is the relation between the Higgs potential parameters and Higgs mass spectra. Once the Higgs potential parameters are determined by the strong dynamics, one can calculate the properties of the Higgs particles just like in the E2HDM. In particular, we remark that all the three neutral Higgs states are mixed and all have non-zero Higgs-gauge-gauge type interactions if the Higgs potential violates CP symmetry, while only two neutral Higgs fields can couple with gauge-gauge state in the CPC case. Therefore, in case we will observe three neutral scalar particles decaying into $W^+W^-$ and/or $ZZ$ states at collider experiments, this would be a smoking gun signature of any 2HDM with CPV. Detailed collider studies of such Higgs decays were performed in the context of the E2HDM with CPV  in Ref.~\cite{Keus:2015hva}. Here, we will  estimate the Branching Ratios (BRs) of these diboson modes in our scenario in Sect.~\ref{sec:Signatures}.

In contrast, there are two unique aspects of the C2HDM.  
Firstly, CPV effects in the Higgs potential correlate with those in the top Yukawa coupling, because the parameters of the former are generated through 
top quark loop contributions containing the latter. 
(This is a rather generic aspect in CHMs with CPV.) 
Secondly, the C2HDM with CPV generally predicts spontaneous CPV, contrarily to the E2HDM case, where both Higgs VEVs can be taken to be real and positive (without loss of generality) by rephasing the Higgs fields and redefining the potential parameters, which is not allowed in our case, so that the relative phase of the two VEVs, $\theta_v$, must be retained. 
In the $SO(6)\to SO(4)\times SO(2)$ CHM,
the complex phase of the Higgs VEVs predicts a non-zero $\hat{T}$-parameter at tree level as follows: 
\begin{align}
\hat{T}
\simeq
-\xi\,\frac{8\tan^2\beta}{(1+\tan^2\beta)^2}\sin^2\theta_{v},  \label{eq:tpara}
\end{align}
where $\xi=v^2_{\rm{EW}}/f^2$.
The bound coming from the EW precision tests is roughly estimated as $|\hat{T}|\lesssim 10^{-3}$.
Therefore, in our setup, $\tan\beta \gg 1$ or $\ll 1$ is required to avoid such a severe constraint on $\hat{T}$, while allowing for sizable CPV in the Higgs sector. 

In order to see the correlation of the CPV effects between the Yukawa sector and Higgs potential, 
let us focus on the top Yukawa sector. The Higgs boson couplings with the top quarks can be read from Eq.~(\ref{eq:Leff}). Up to $\mathcal{O}(\Phi^5_{1,2})$, we obtain
\begin{align}
-\mathcal{L}_{\rm{eff}}
\ni
-i\,
\bar{q}_L (
Y_{t,1}i\sigma_2 \Phi^*_1
+
Y_{t,2}i\sigma_2 \Phi^*_2
)t_R
+
h.c.
\,,
\end{align}
where $Y_{t,1}$ and $Y_{t,2}$ are calculated as
\begin{align}
Y_{t,1}
&=
\biggl(
\frac{M_2}{f}
-\frac{4}{3}\frac{M_1}{f^3}\Phi^\dag_1\Phi_1
+\frac{1}{3}\frac{M_1}{f^3}(\Phi^\dag_1\Phi_2+\Phi^\dag_2\Phi_1)
\biggr)
\cos\theta_t
\nn\\
&
~~+
i\biggl(
\frac{M_1}{f}
-\frac{1}{3}\frac{M_1}{f^3}\Phi^\dag_1\Phi_1
+\frac{2}{3}\frac{M_2}{f^3}(\Phi^\dag_1\Phi_2+\Phi^\dag_2\Phi_1)
\biggr)
\sin\theta_t
\,,\\
Y_{t,2}
&=
\biggl(
\frac{M_1}{f}
-\frac{M_1}{f^3}\Phi^\dag_1\Phi_1
-\frac{1}{3}\frac{M_1}{f^3}\Phi^\dag_2\Phi_2
-\frac{2}{3}\frac{M_2}{f^3}(\Phi^\dag_1\Phi_2+\Phi^\dag_2\Phi_1)
\biggr)
\cos\theta_t
\nn\\
&
~~+
i\biggl(
\frac{M_2}{f}
-\frac{4}{3}\frac{M_2}{f^3}\Phi^\dag_2\Phi_2
-\frac{1}{3}\frac{M_1}{f^3}(\Phi^\dag_1\Phi_2+\Phi^\dag_2\Phi_1)
\biggr)
\sin\theta_t
\,,
\end{align}
with the form factors being evaluated in the zero momentum limit. 
The top mass is then extracted as 
\begin{align}
m_t = -i\frac{v}{\sqrt{2}}[Y_{t,1}|_{\langle\Phi_i \rangle}\cos\beta +  Y_{t,2}|_{\langle\Phi_i \rangle}\sin\beta e^{-i\theta_v} ]\,,
\end{align}
where $Y_t|_{\langle\Phi_i \rangle}$ denotes $Y_t$ evaluated by the replacement of $\Phi_i$ by $\langle\Phi_i \rangle$.
Therefore, 
it should be noted that, in the C2HDM, the top mass and Yukawa coupling depend on $\theta_t$ and $M_{1,2}(0)$,  which also appear in the Higgs potential. This is indeed
the correlation between the top and Higgs sectors in the C2HDM that we mentioned in the previous section.

\section{Signatures of the C2HDM with CPV}
\label{sec:Signatures}

In this section, we show some numerical results in the C2HDM, particularly focusing on the dependence upon CPV phases of various quantities such as couplings of the SM-like Higgs boson and decays of additional Higgs bosons. 
For concreteness, we consider the case with two generations of $\Psi_t$ fields ($N = 2$) 
and assume that
\begin{align}
&g_\rho=g_{\rho_X}\,,\\
&f_1=f_2=\sqrt{2}f\,,\\
&\Delta^2_L
=
\Delta^1_R
=
M^{21}_\Psi
=
Y^{11}_1
=
Y^{22}_1
=
Y^{11}_2
=
Y^{22}_2
=
Y^{21}_1
=
Y^{21}_2
=
0\,.
\label{eq:LRsym}
\end{align}
We note that the UV finiteness in the momentum integration for the form factors is  automatically ensured with these assumptions. (See Appendix~\ref{app:Higgspotential} for the details.)
In this setup, the input parameters are listed as follows:
\begin{align}
f,~g_\rho,~
\Delta^1_L,~
\Delta^2_R,~
M^{11}_\Psi,~
M^{22}_\Psi,~
M^{12}_\Psi,~
Y^{12}_1,~
Y^{12}_2,
~\theta_t. 
\end{align}
Without  loss of generality, we can take $\Delta^1_L$,~$\Delta^2_R$,~$M^{11}_\Psi$,~$M^{22}_\Psi$, $M^{12}_\Psi$ and $\bar{Y}^{12}_2 \equiv M^{12}_\Psi - Y^{12}_2$ to be positive and real by rephasing 
the fields in the strong sector, while $Y^{12}_1$ is complex. 
In the following analysis, we assume $Y^{12}_1$ to be real, so that the effect of CPV solely comes from the $\theta_t$ parameter. 
In this setup, the top mass is expressed as 
\begin{align}
m_t &\equiv  |m_t| = \frac{v_{\rm SM}}{\sqrt{2}}Z_t\Bigg[\left\{\left(\frac{Y_2^{12}}{f}\cos\beta + \frac{Y_1^{12}}{f}\sin\beta \cos\theta_v \right)\cos\theta_t + \frac{Y_2^{12}}{f}\sin\theta_t\sin\beta \sin\theta_v\right\}^2  \notag\\
&+ \left\{\left(\frac{Y_1^{12}}{f}\cos\beta + \frac{Y_2^{12}}{f}\sin\beta \cos\theta_v \right)\sin\theta_t -\frac{Y_1^{12}}{f}\sin\beta \cos\theta_t \sin\theta_v \right\}^2 \Bigg]^{1/2} + {\cal O}(\xi^{3/2}), \label{eq:mt2}
\end{align}
where 
\begin{align}
Z_t = \frac{\Delta_L^1 \Delta_R^2}{M_\Psi^{11}M_\Psi^{22}}
\left[1 + (\Delta_L^1)^2\frac{(M_{\Psi}^{12})^2 + (M_{\Psi}^{22})^2}{(M_{\Psi}^{11})^2 (M_{\Psi}^{22})^2}  \right]^{-1/2}
\left[1 + (\Delta_R^2)^2\frac{(Y_1^{12})^2 + (\bar{Y}_2^{12})^2}{(M_{\Psi}^{11})^2 (M_{\Psi}^{22})^2}  \right]^{-1/2}. 
\end{align}
In the numerical evaluation, we take into account the ${\cal O}(\xi^{3/2})$ contribution to the top mass in order to consistently 
include the ${\cal O}(\xi)$ corrections in the modified top Yukawa coupling. 

Let us now briefly explain the strategy of our numerical analysis. 
First, we scan the parameters in the following ranges:  
\begin{align}
&1 \leq g_\rho \leq 5, \quad 0 \leq \theta_t \leq \pi, \notag\\
&f \leq \{M^{11}_\Psi,\Delta^1_L,\Delta^2_R,M^{12}_\Psi\}  \leq 5f,\quad 
-5f \leq \{Y^{12}_1,Y^{12}_2\}  \leq 5f, \label{scan}
\end{align}
with $\bar{Y}^{12}_2 \geq 0$ and $M^{22}_\Psi = M^{11}_\Psi$. 
We note that the case with $-\pi < \theta_t < 0$ shows symmetric results with respect to those with $0 < \theta_t < \pi$, because the potential parameters and top mass are 
given as functions of $2\theta_t$. We thus focus on the positive values of $\theta_t$. 
For each point, we evaluate all the form factors given in Appendix~\ref{eq:formfactors} and, 
by performing the integration on the loop momentum, we obtain numerical values for the parameters of the Higgs potential, i.e.,  $m_i^2$ ($i=1,\dots,3$) and $\lambda_j$ ($j=1,\dots,7$),
from the formulae given in Appendix~\ref{app:Higgspotential}. 
Next, we numerically solve the tadpole conditions
\begin{align}
\frac{\partial V}{\partial \tilde{\phi}_i^0}\Bigg|_0 = 0,\quad (i=1,2,3), \label{eq:tadpole}
\end{align}
where $X|_0$ denotes the value of $X$ in the limit of zero field values while 
$\tilde{\phi}^0_{1,2,3}$ are the neutral components of the Higgs fields, see Eq.~(\ref{eq:components}). 
These equations are given in terms of the potential parameters, the VEVs ($v_1,v_2$) and their relative phase $\theta_v$. 
In the E2DHM, these equations can  be solved analytically, so that some of the dimensionful parameters of the potential, e.g., $m_1^2$, $m_2^2$ and $\text{Im} (m_3^2)$, 
can be written in terms of the others and the VEVs. In contrast,  in the C2HDM, 
we cannot analytically solve them.  Instead, we numerically find a set of values ($v_1,v_2,\theta_v$) which satisfy Eq.~(\ref{eq:tadpole}), amongst which 
we choose the combination giving the deepest potential value if several extrema appear at the same time. 
Finally, we evaluate various physical quantities (such as masses of the Higgs bosons, their couplings and decay BRs) in a similar way to that in the E2HDM (as intimated). 
In the last step, we impose the following constraints: 
\begin{align}
|\hat{T}|<10^{-3} \label{eq:c_tpara}
\end{align}
and 
\begin{align}
10\times |v_{\rm SM}^{} - 246~\text{GeV}| + 
|m_t - 173.3~\text{GeV}| + 
|m_{H_1}^{} - 125.1~\text{GeV}| \leq 20~{\rm GeV}, \label{eq:c_masses}
\end{align}
where $m_{H_1}^{}$ is the mass of the lightest neutral Higgs boson. 
For the VEV, we impose a much tighter constraint as compared with those on the other two observables, as 
it is most accurately measured by experiments.  In the following analysis, we fix $f = 1$ TeV as a reference value. 

\begin{figure}[t]
\begin{center}
\includegraphics[width=70mm]{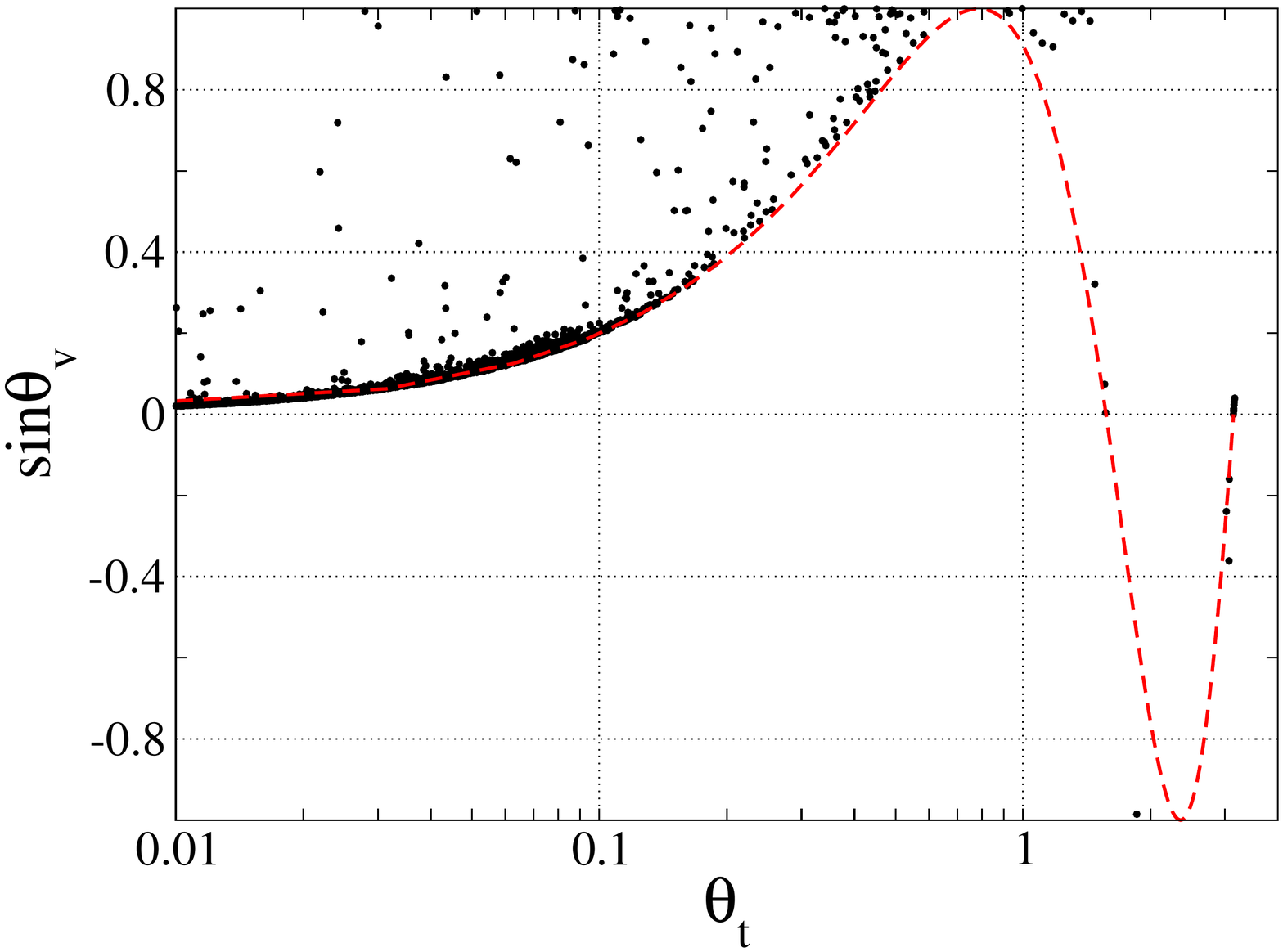}
\includegraphics[width=70mm]{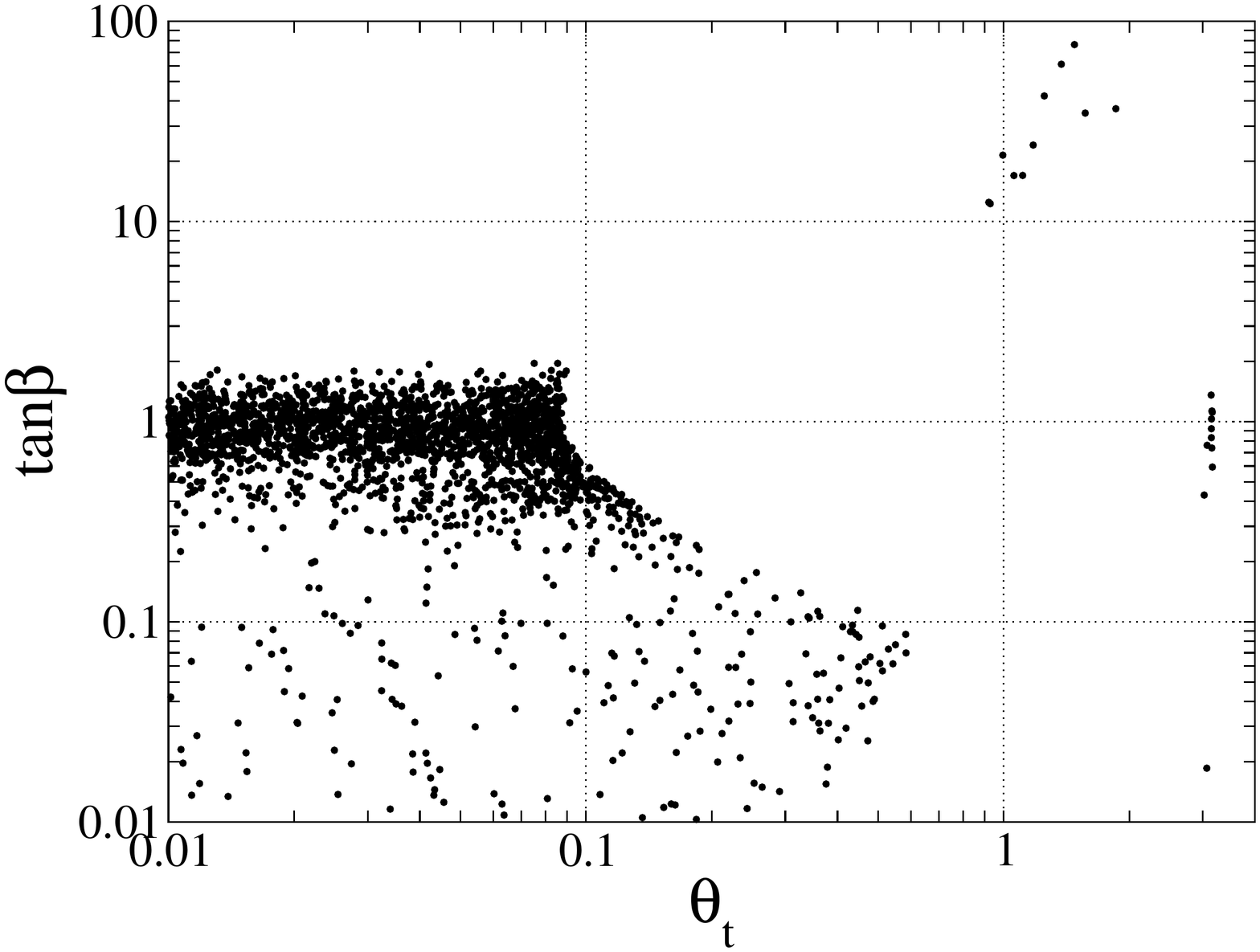}
\end{center}
\caption{Correlations between $\theta_t$ and $\sin\theta_v$ (left) as well as $\theta_t$ and $\tan\beta$ (right). In the left plot, the red dashed curve represents $\sin 2\theta_t$. 
}
\label{fig:tanb}
\end{figure}

First of all, in Fig.~\ref{fig:tanb}, we show the behaviour of $\sin\theta_v$ and $\tan\beta$ as functions of $\theta_t$. 
In this figure (and in all the following ones), black points are allowed by the constraints from Eqs.~(\ref{eq:c_tpara}) and (\ref{eq:c_masses}). 
In addition, we show the case with $\theta_t \geq 0.01$, because that with $\theta_t < 0.01$ is almost the same as the CPC scenario. 
In the left plot, we see that the allowed points mostly appear on the curve of $\sin 2\theta_t$, which comes from the fact that the imaginary part of the potential parameters 
is given as a function of $\sin 2\theta_t$, see Eq.~(\ref{eq:ff-special}). 
We note that most of the allowed points are numerically found to have $\theta_v \sim \pi - 2\theta_t$. 
There appear a few points above the dashed curve, which typically require $\tan\beta \ll 1$ due to the constraint from the $\hat{T}$ parameter, see Eq.~(\ref{eq:tpara}),  
as it can be clearly seen by looking at the density of the points between the left and right plots of Fig.~\ref{fig:tanb}. 
In the right plot of this figure, we see the $\theta_t$ dependence of $\tan\beta$. When the CPV phase $\theta_t$ is negligibly small, only the upper limit of $\tan\beta \sim 1$ appears from the condition of Eq.~(\ref{eq:c_masses})\footnote{We checked that, if we take a larger scan region of the strong sector parameters, e.g., up to $10 f$ instead of $5f$ in Eq.~(\ref{scan}), the maximal value of $\tan\beta$ at $\theta_t \sim 0$ is extended to be about 10, 
which is consistent with  previous results~\cite{DeCurtis:2018zvh}.  }.
For $\theta_t\gtrsim 0.1$, the upper limit of $\tan\beta$ is getting more severe, because of the constraint from the $\hat{T}$ parameter. 
For $\theta_t \sim \pi/2$, points are distributed over a larger $\tan\beta$ region. This is again due to the constraint from the $\hat{T}$ parameter, but in this region 
the role of $\sin\beta$ and $\cos\beta$ in the expression of $m_t$ given in  Eq.~(\ref{eq:mt2}) 
is interchanged as compared with the case with $\theta_t \ll 1$, so that large $\tan\beta$ values are favoured over small ones. 

\begin{figure}[t]
\begin{center}
\includegraphics[width=50mm]{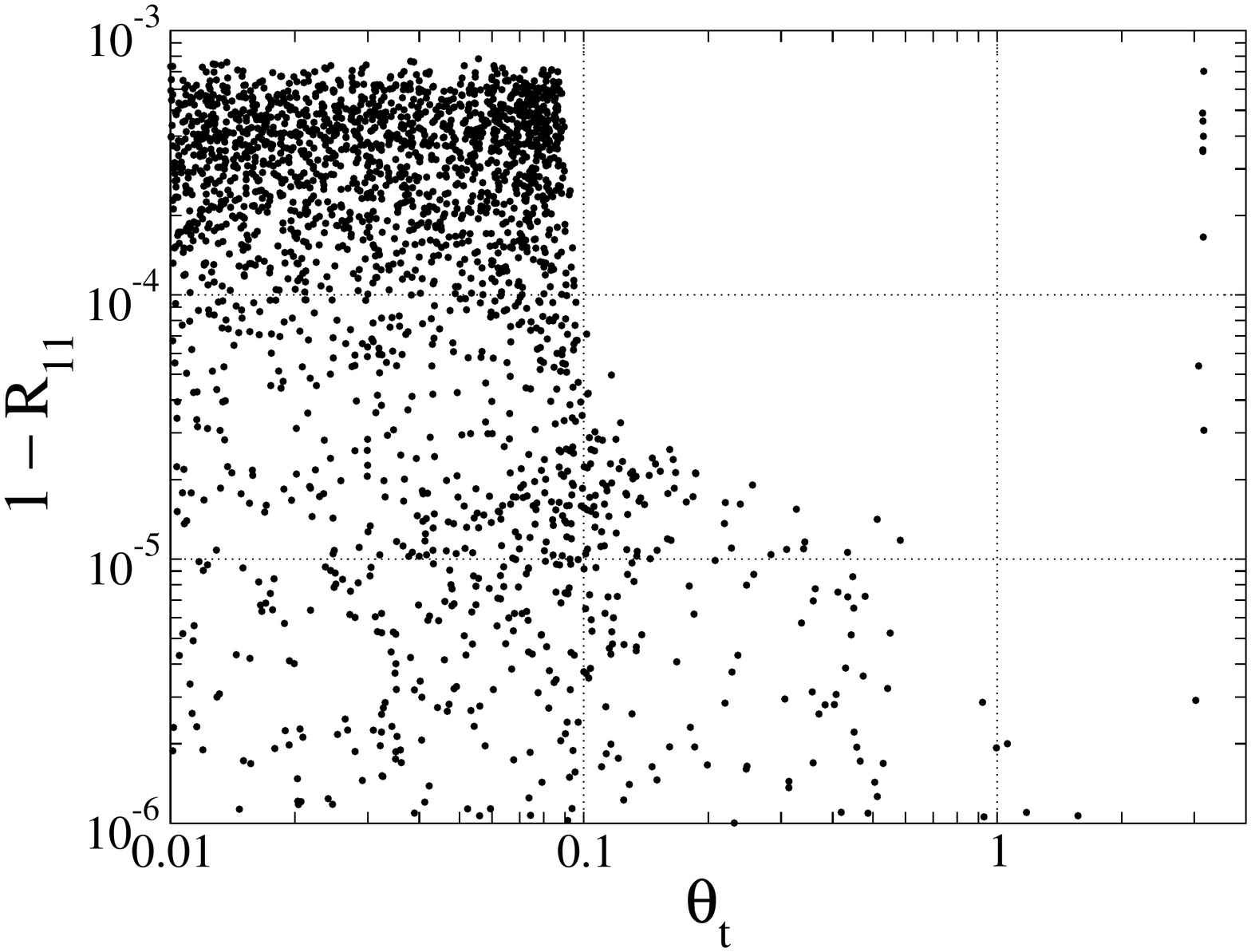}
\includegraphics[width=50mm]{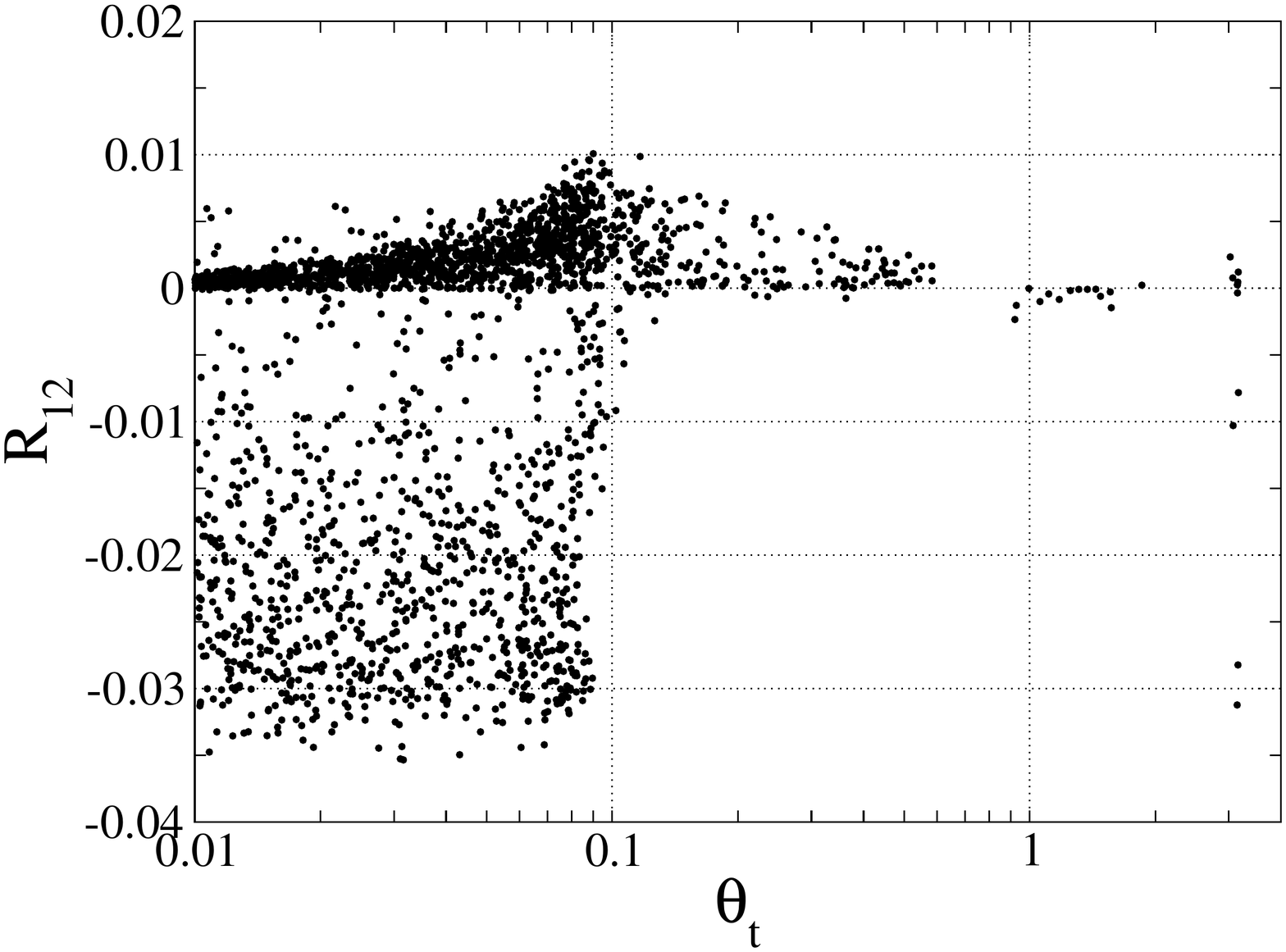}
\includegraphics[width=50mm]{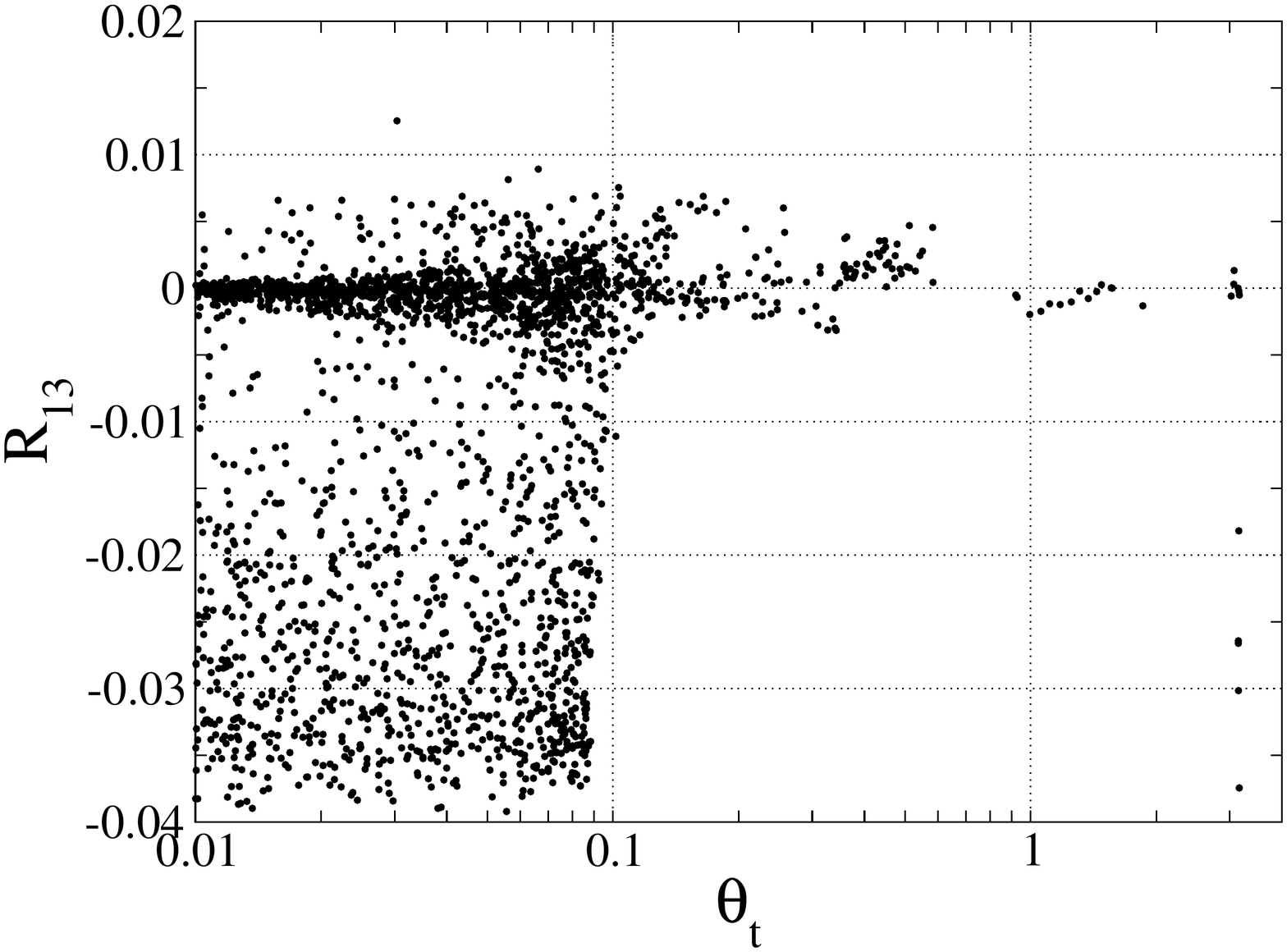}
\end{center}
\caption{$\theta_t$ dependence of the values of $1 - R_{11}$ (left), $R_{12}$ (center) and $R_{13}$ (right), where the mixing matrix $R$ is defined in Eq.~(\ref{eq:rmat}). }
\label{fig:rmat}
\end{figure}

\begin{figure}[t]
\begin{center}
\includegraphics[width=70mm]{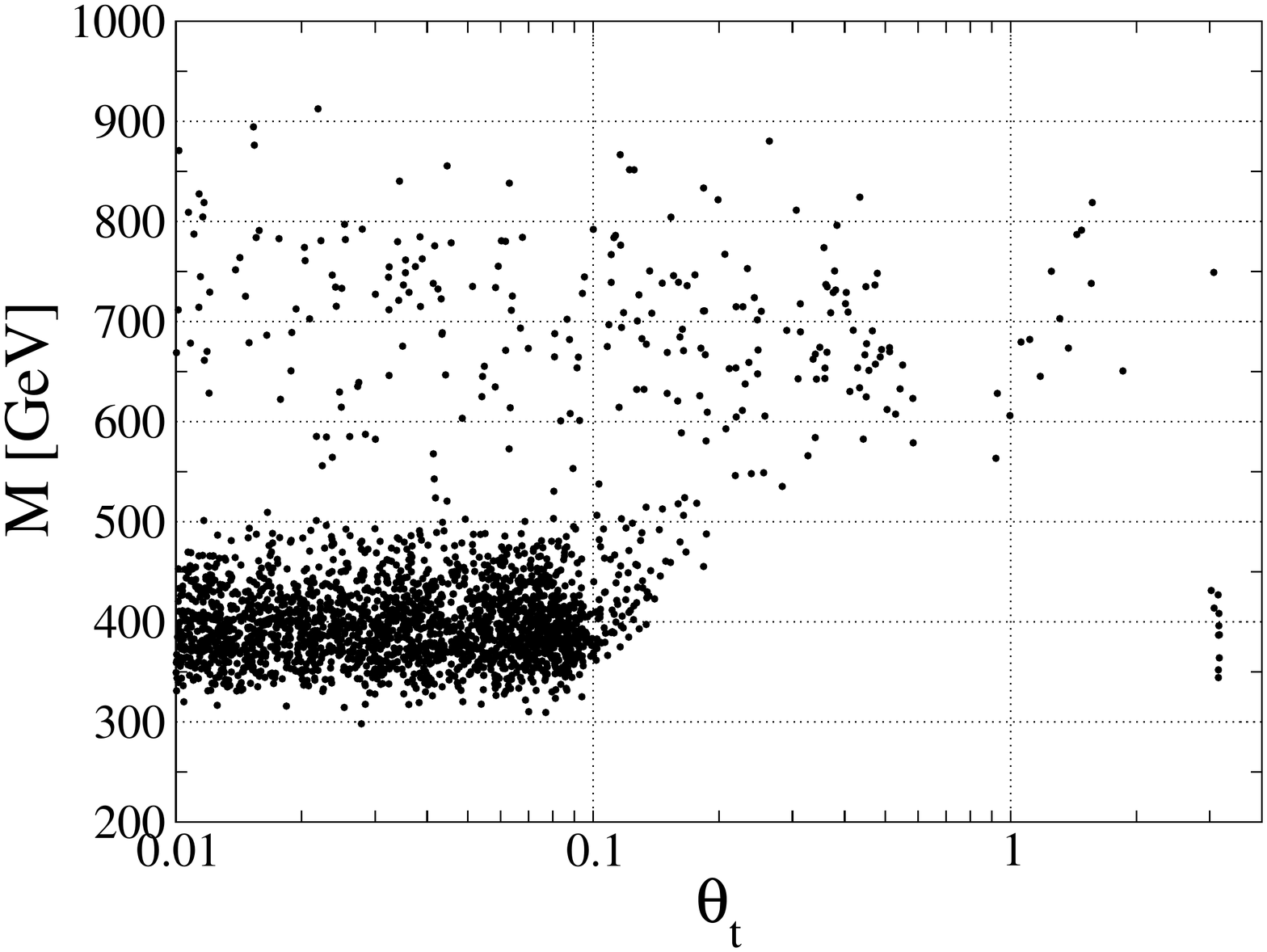}
\includegraphics[width=70mm]{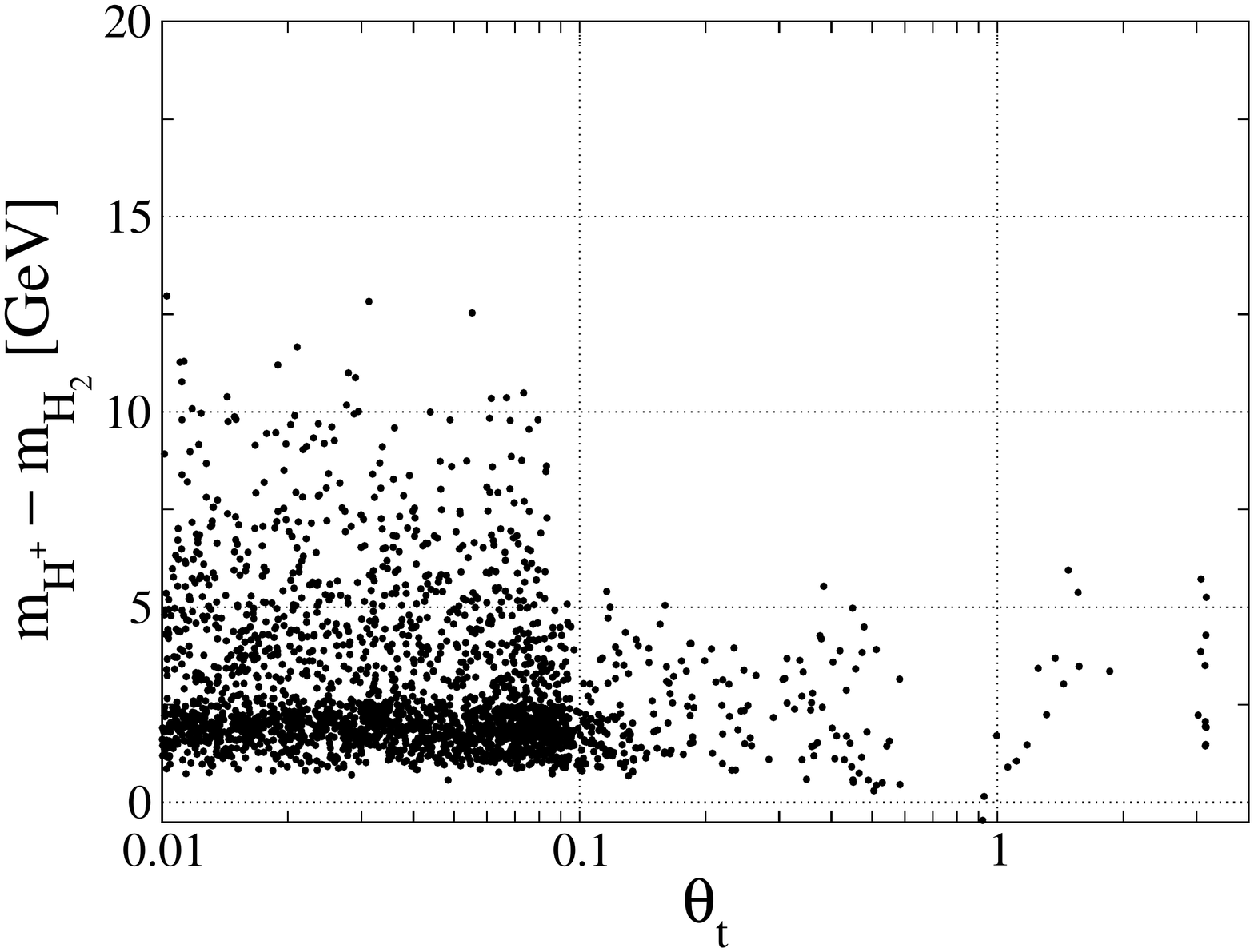}\\
\includegraphics[width=70mm]{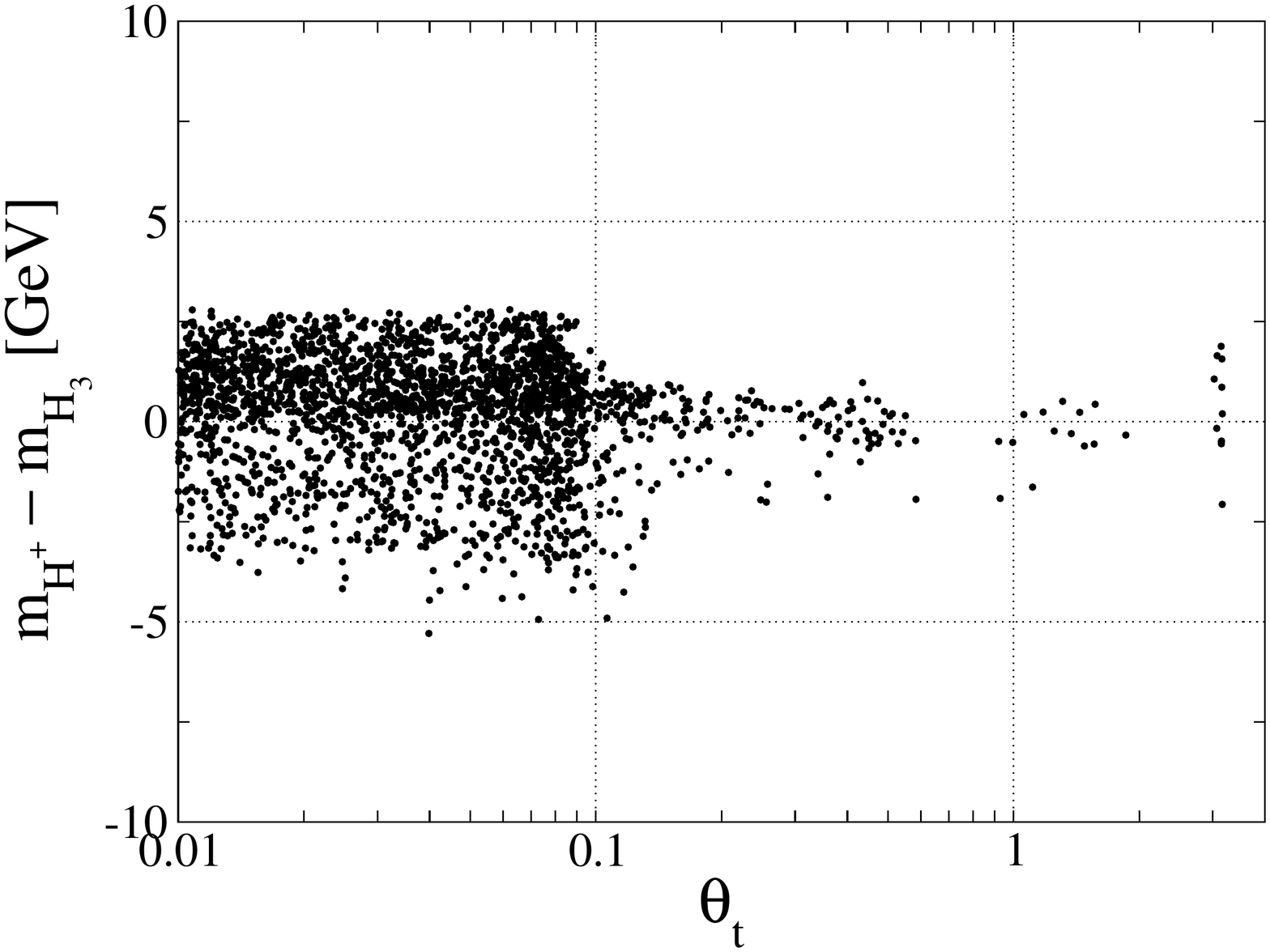}
\includegraphics[width=70mm]{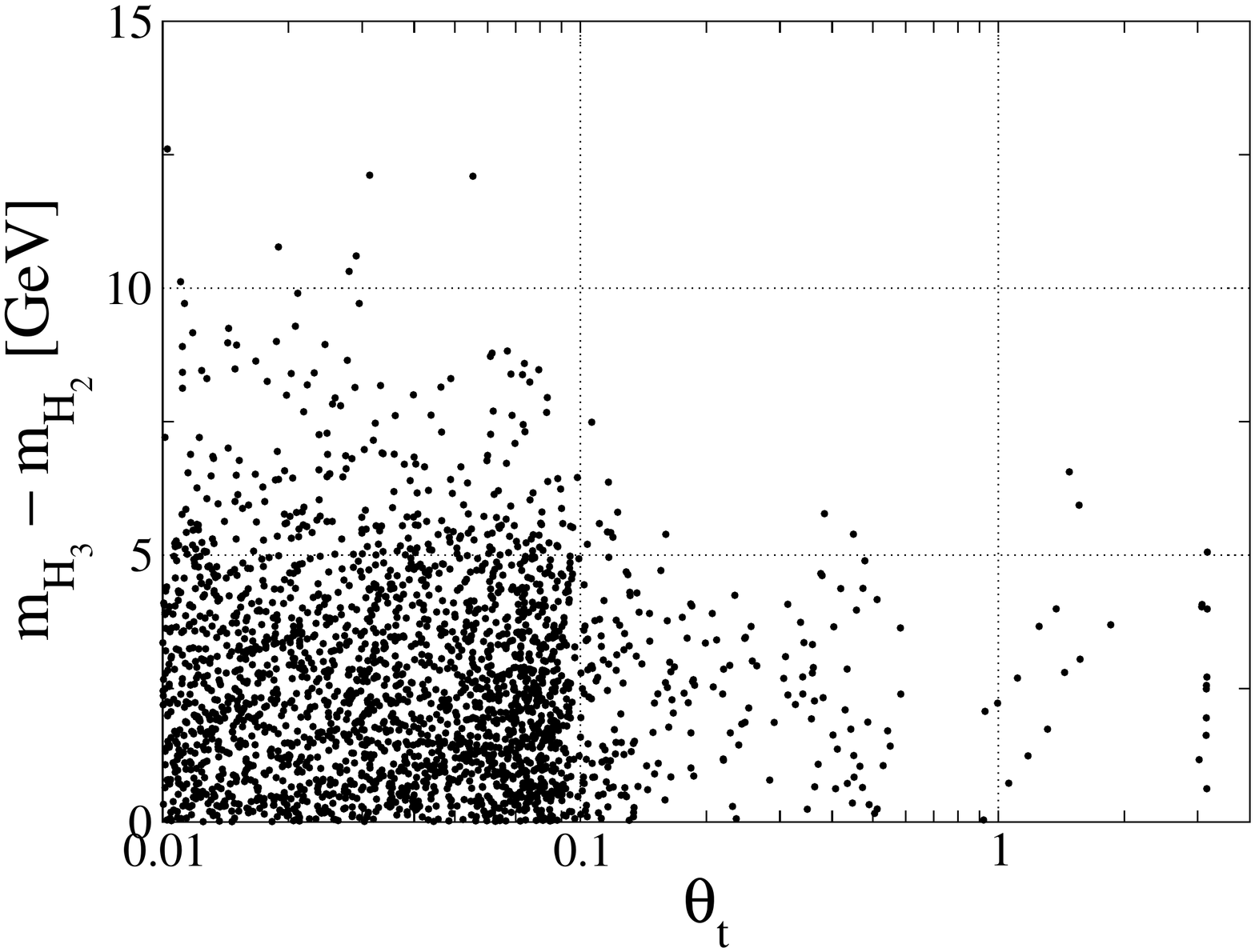}
\end{center}
\caption{Dimensionful parameter $M$ (top-left) and the mass differences between two additional Higgs bosons as functions of $\theta_t$. 
}
\label{fig:masses}
\end{figure}

Next, in Fig.~\ref{fig:rmat}, we discuss the behaviour of the elements of the mixing matrix $R_{ij}$ defined in Eq.~(\ref{eq:rmat}). 
In particular, the value of $1 - R_{11}$ describes the ``alignmentness'' of the 125 GeV Higgs boson, i.e., 
in the limit of $1 - R_{11} \to 0$, the $H_1$ state coincides with the $\tilde{\phi}_1^0$ state in the Higgs basis. 
In this case, deviations in the $H_1$ couplings to ordinary particles from their SM predictions only come from the parameter 
$\xi$, which describes the compositeness of the Higgs boson. This should be compared with the E2HDM case, where the Higgs boson couplings 
become the same as the ones of the SM Higgs boson at tree level in the alignment limit. 
As seen from the left plot of Fig.~\ref{fig:rmat}, our scenario is almost the alignment limit, as the deviations from it are  quite small, i.e., ${\cal O}(10^{-3})$ or smaller for $\theta_t \lesssim 0.1$. 
For larger values of $\theta_t$, alignmentness becomes stronger, because the masses of the heavier Higgs bosons become larger, see Fig.~\ref{fig:masses}, and thus 
the decoupling behaviour turns out to be very strong. 
We also show the different elements of $R$ in the center and right plots. It is seen that the magnitude of $R_{12}$ and $R_{13}$ can maximally be ${\cal O}(0.01)$. 
This small mixing plays an important role to get sizable BRs of $H_2 \to VV$ and $H_3 \to VV$ ($V = W^\pm,Z$), as we will see below. 

\begin{figure}[t]
\begin{center}
\includegraphics[width=75mm]{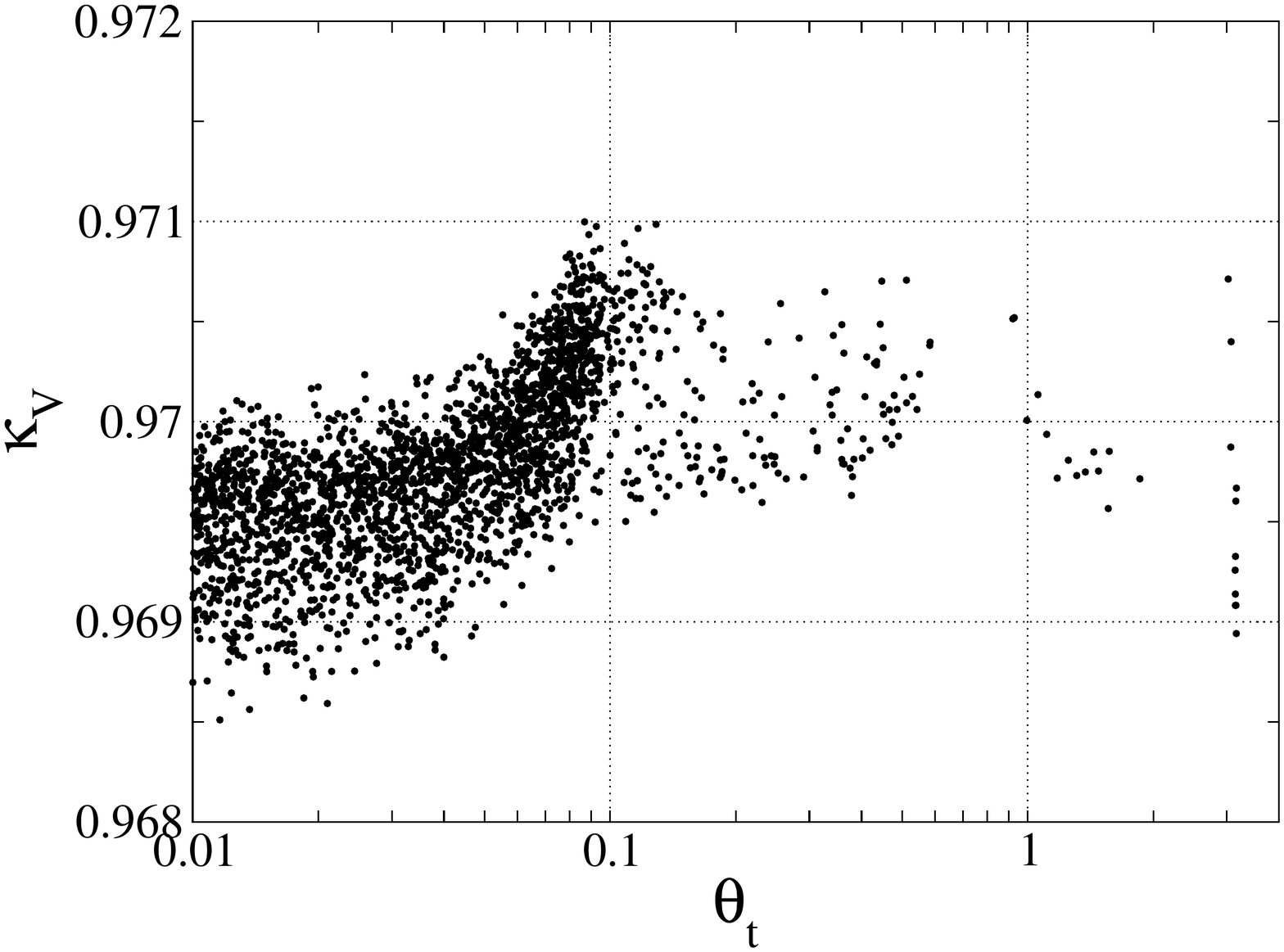} 
\includegraphics[width=75mm]{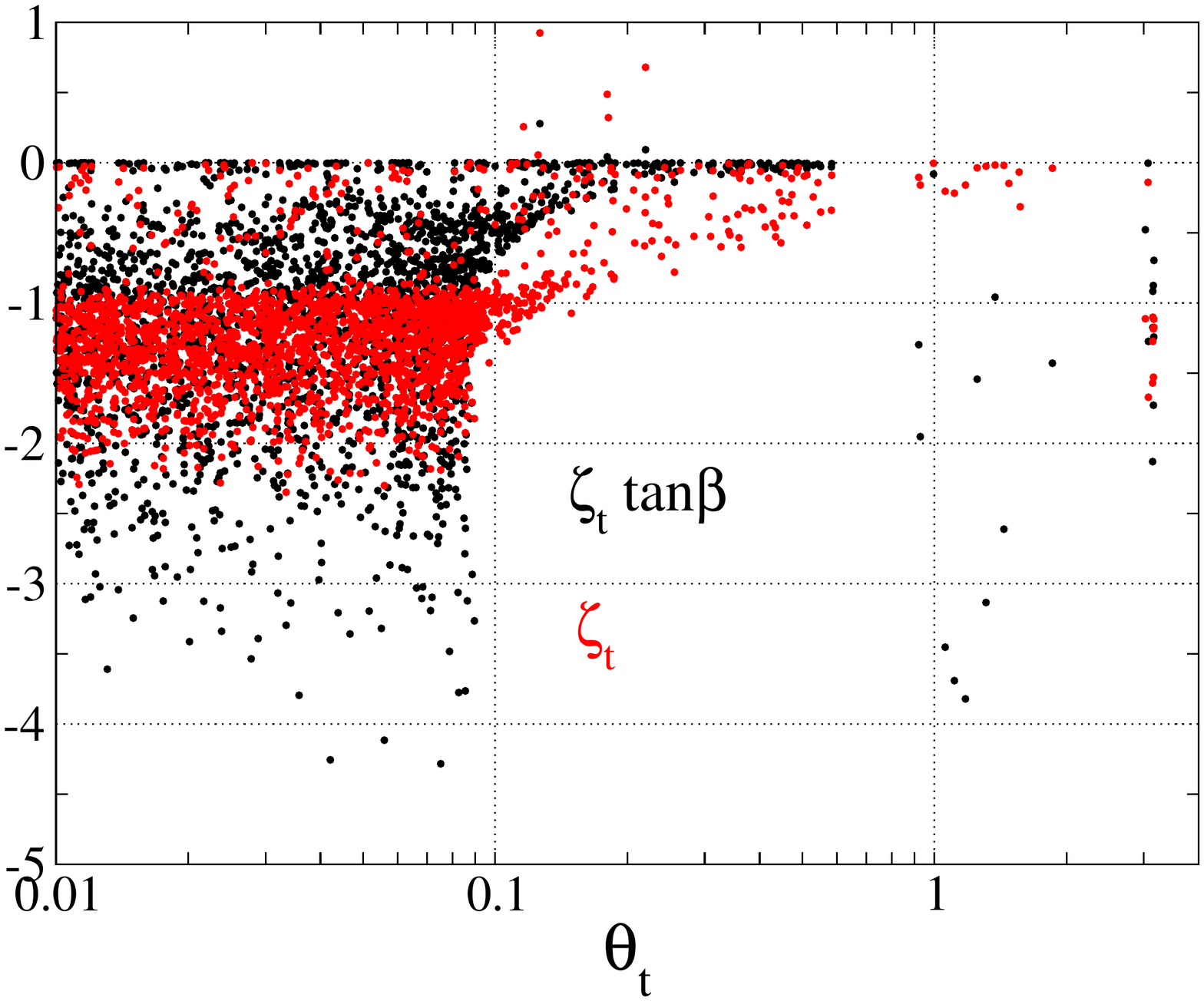}\\
\includegraphics[width=75mm]{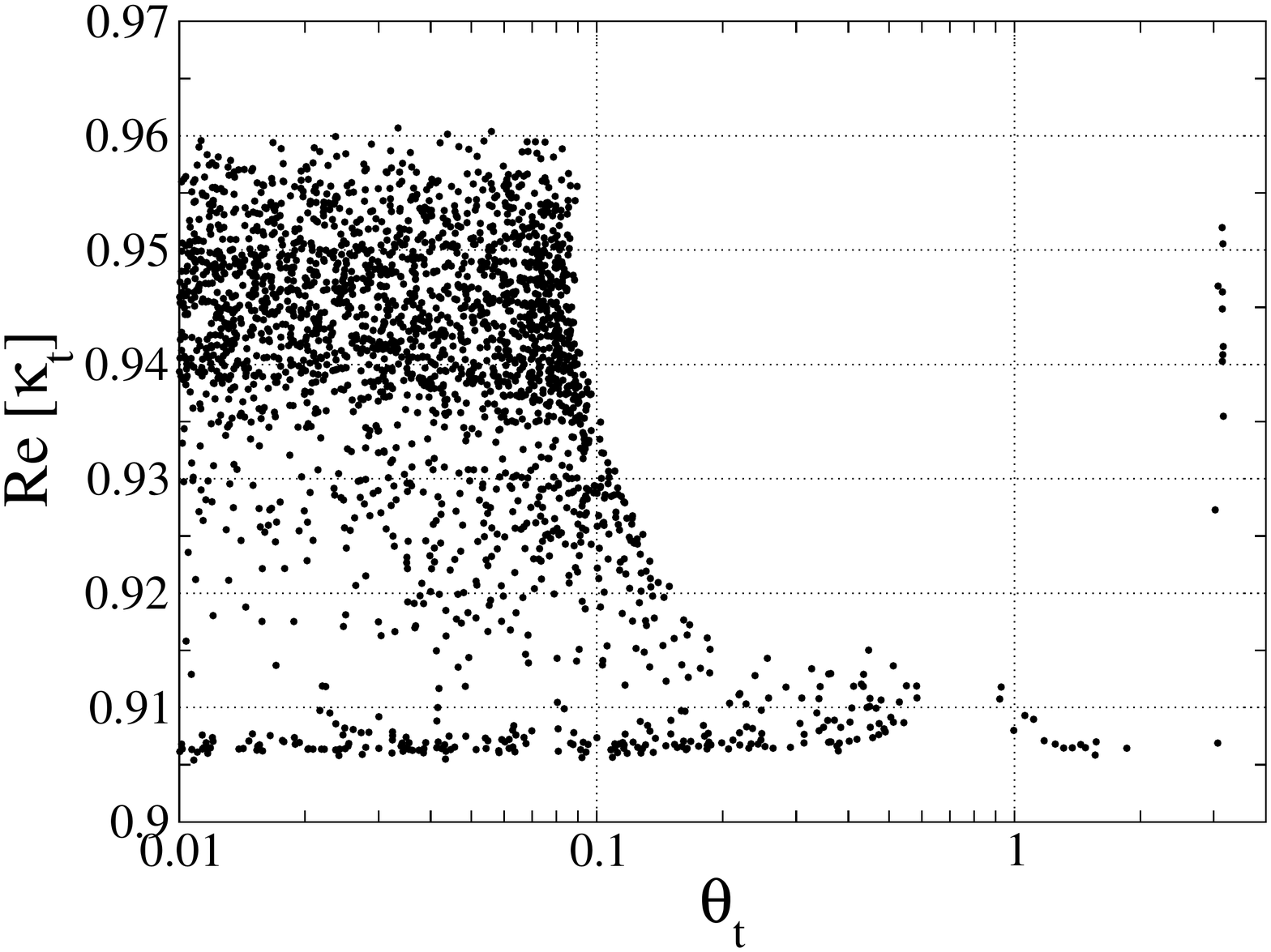}
\includegraphics[width=75mm]{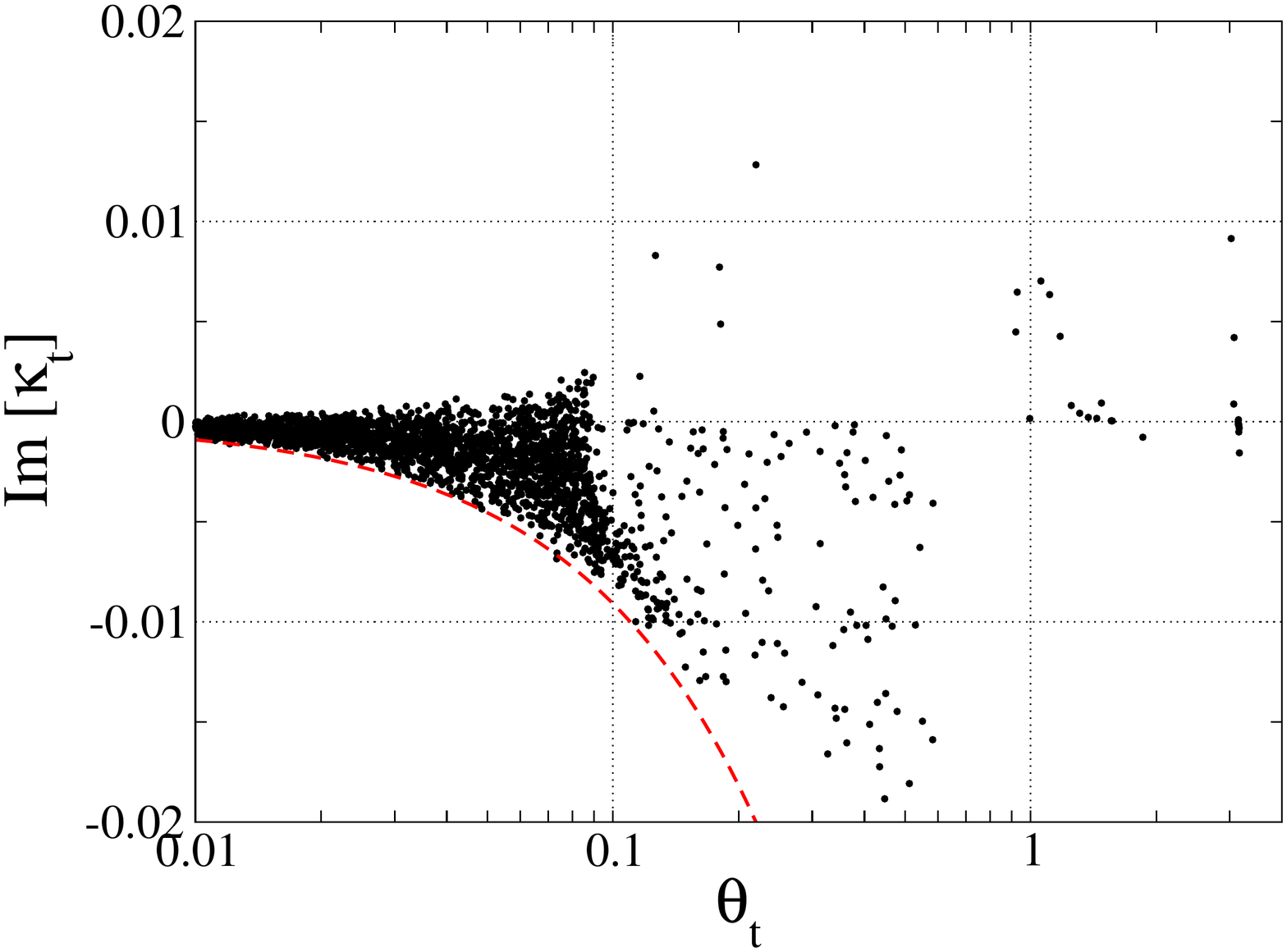}
\end{center}
\caption{Scaling factors $\kappa_V^{}$ (top-left),  Re$[\kappa_t^{}]$ (lower-left) and Im$[\kappa_t^{}]$ (lower-right) as functions of $\theta_t$. 
The values of $\zeta_t$ and $\zeta_t \tan\beta$ are also shown (top-right). 
The red dashed curve shows the value of $\xi \theta_t \zeta_t$ with $\zeta_t = -1.5$. }
\label{fig:kappa}
\end{figure}

From these results, we expect that the deviations in the Higgs boson couplings can be well approximated in the alignment limit by analytical formulae. 
Let us define the usual scaling factors $\kappa_X$ as ($X=V$ and $t$, where $V=W^\pm,Z$)
\begin{align}
\kappa_X = \frac{g_{hXX}^{\rm C2HDM}}{g_{hXX}^{\rm SM}}. 
\end{align}
In the alignment limit, $\kappa_X$ can be expressed by using the relation $\theta_v = \pi - 2\theta_t$ as 
\begin{align}
\kappa_V &=1-\frac{\xi}{2}\l(1-\frac{1}{2}\sin^2 2\beta \sin^22\theta_t\r) +\mathcal{O}(\xi^2), \label{eq:kappav} \\
\text{Re}[\kappa_t]&= 1 - \left(\frac{3}{2} + \frac{\zeta_t\tan\beta }{1 - \zeta_t\tan\beta}\right)\xi + {\cal O}(\xi^2, \theta_{t}^2), \\
\text{Im}[\kappa_t]&= \frac{\xi \theta_t }{6(1 - \zeta_t\tan \beta)^2}\Bigg\{6\zeta_t(1 - \tan^2\beta) + [\zeta_t^2 - 8  - (8-7\zeta_t +4\zeta_t^2)\sin 2\beta]\tan\beta \notag\\
& - (8 + 4\zeta_t + \zeta_t^2)\tan\beta \cos2\beta \Bigg\}+ {\cal O}(\xi^2, \theta_{t}^3),  \label{eq:kappat} 
\end{align}
where $\zeta_t \equiv  Y_1^{12}/Y_2^{12}$. 
In Fig.~\ref{fig:kappa}, we show the predicted $\kappa_X^{}$ values as functions of $\theta_t$. We confirm that these values can be well explained by using the approximate formulae given in the above equations. 
In the CPC limit $\theta_t \to 0$ with $f = 1$ TeV, $\kappa_V^{}$ is calculated to be about 0.97 using $\kappa_V \sim  1 - \xi/2$. 
In fact, most of the points appear at $\kappa_V \lesssim 0.97$ and $\theta_t \simeq 0$ with a small fluctuation due to tiny mixing effects, i.e., $R \neq I$. 
It is seen that $\kappa_V$ is slightly enhanced at $\theta_t \gtrsim 0.1$, because the phase $\theta_t$ slightly cancels the negative correction proportional to $\xi$, see Eq.~(\ref{eq:kappav}). 
For Re$[\kappa_t^{}]$, in the CPC limit, the lowest (largest) value $\sim 0.91~(0.96)$ can be explained with $\zeta_t \tan\beta \sim 0 \, (-4)$, see also the top-right panel of Fig.~\ref{fig:kappa}. 
When $\theta_t$ gets larger, the value of Re$[\kappa_t^{}]$ rapidly shrinks to its lowest value $\sim 0.91$, 
because $\zeta_t \tan\beta$ approaches zero due to the fact that only small values of $\tan\beta$ are allowed by the $\hat{T}$ parameter, see Fig.~\ref{fig:tanb}. 
Regarding Im$[\kappa_t^{}]$, its magnitude gets larger when $\theta_t$ increases, e.g., $|\text{Im}[\kappa_t^{}]| \lesssim 0.01$ at $\theta_t \simeq 0.1$. 
The maximal value can be well fit by the dashed curve which is given by the leading term of Eq.~(\ref{eq:kappat}), i.e., $\xi \theta_t \zeta_t$ with $\zeta_t = -1.5$. 
In addition, it is important to mention here that the effect of a non-zero $\theta_t$ on $\kappa_V$ and Re$[\kappa_t]$ is anti-correlated. 

Next, we discuss the mass spectrum of the heavy Higgs bosons and their phenomenology. 
The masses of heavier Higgs states are expressed in the limit of Im$[\lambda_5]=0$ and $R = I$ as
\begin{align}
m_{H^\pm}^2 & =  M^2 - \frac{v^2}{2}\left(\lambda_4 + \text{Re}[\lambda_5] + \frac{\cot\beta}{\cos\theta_v}\text{Re}[\lambda_6] + \frac{\tan\beta}{\cos\theta_v}\text{Re}[\lambda_7]\right), \\
m_{H_2}^2  & = \text{Min}[({\cal M}_N)_{22},({\cal M}_N)_{33}],\quad m_{H_3}^2 = \text{Max}[({\cal M}_N)_{22},({\cal M}_N)_{33}], 
\end{align}
where $({\cal M}_N)_{ij}$ are the mass matrix elements of the neutral Higgs bosons in the Higgs basis defined in Eq.~(\ref{hessian}). The (2,2) and (3,3) elements are given by 
\begin{align}
({\cal M}_N)_{22} &=M^2 + \frac{v^2}{4}\Big[\sin{2\beta}(\lambda_1 + \lambda_2 -2 \lambda_3 -2\lambda_4)-2(1 - \cos^2{2\beta}\cos{2\theta})\text{Re}[\lambda_5] \notag\\
& -\frac{2}{\cos \theta_v}(\cos^2 \theta_v \sin{4\beta} + \cot\beta)\text{Re}[\lambda_6] + \frac{2}{\cos \theta_v}(\cos^2 \theta_v \sin{4\beta} - \tan\beta)\text{Re}[\lambda_7] \notag\\
& +2\sin{4\beta}\sin\theta(\text{Im}[\lambda_6] - \text{Im} [\lambda_7] )\Big], \\
({\cal M}_N)_{33} & = M^2 - \frac{v^2}{2}\left[2\cos^2 \theta_v \text{Re} [\lambda_5] + \frac{\cot\beta}{\cos \theta_v}\text{Re} [\lambda_6] + \frac{\tan\beta}{\cos \theta_v}\text{Re} [\lambda_7] \right], 
\end{align}
with 
\begin{align}
M^2 \equiv \frac{\text{Re}[m_{3}^2]}{\cos \theta_v}\left(\cot\beta + \tan\beta\right). \label{eq:bmsq}
\end{align}
In our numerical analysis, the $\lambda_i ~(i=1, ... , 7)$ parameters are typically of order 0.1, so that the magnitude of the squared masses is roughly determined by the size of the $M^2$ parameter. 
In Fig.~\ref{fig:masses}, we thus show the typical mass scale $M$ of the heavy Higgs bosons (top-left). 
It is seen that the lowest value of $M$ becomes larger when $\theta_t \gtrsim 0.1$, which can be understood from Eq.~(\ref{eq:bmsq}) with $\tan\beta \ll 1$. 
We can also see that the masses of the additional Higgs bosons are nearly degenerate, i.e., typical mass differences are at the few GeV level. 
This means that the additional Higgs bosons almost do not decay into a lighter additional Higgs boson (or, rather, they do so via a very off-shell gauge boson,
where possible), while they can decay into the SM-like Higgs boson $H_1$. (This is not dissimilar from what found in the C2HDM with CPC.)

\begin{figure}[t]
\begin{center}
\includegraphics[width=75mm]{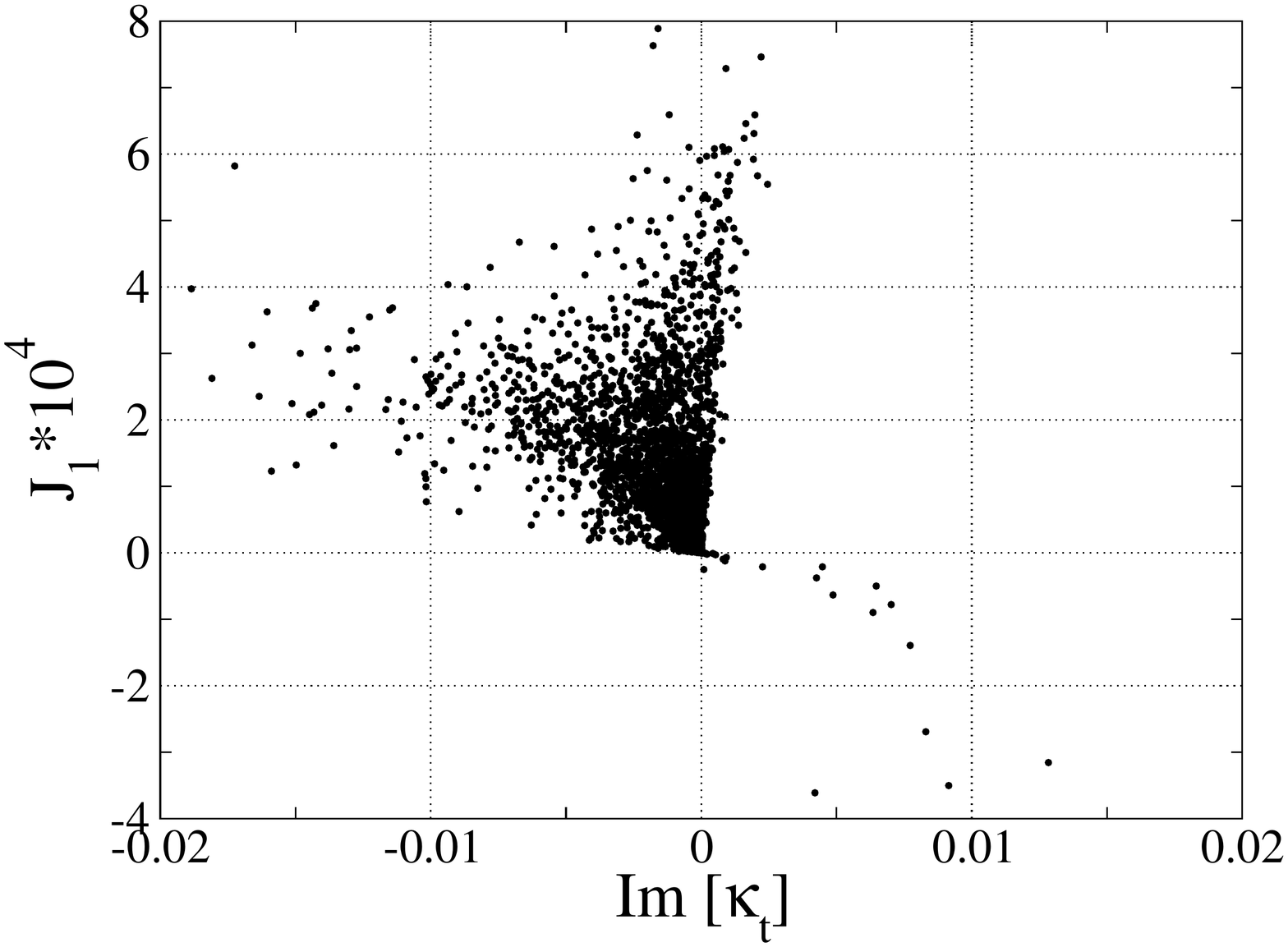}
\includegraphics[width=75mm]{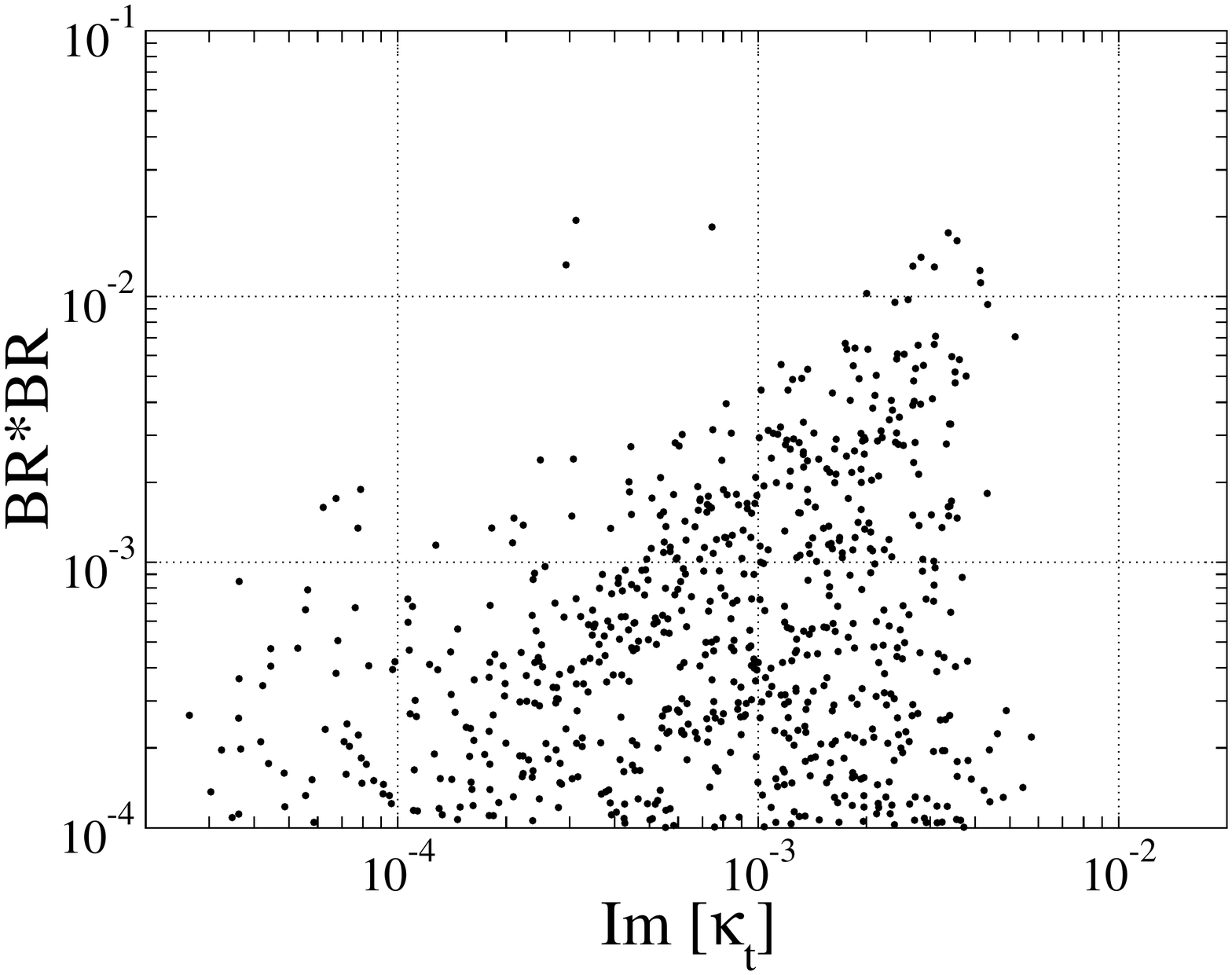}
\end{center}
\caption{Correlation between Im[$\kappa_t$] and the invariant $J_1$ multiplied by $10^4$ (left) as well as  BR($H_2 \to VV$) $\times$ BR($H_3 \to VV$), with $V=W^\pm,Z$ (right). }
\label{brs}
\end{figure}

Finally, we discuss the correlations between CPV effects in the top Yukawa coupling and the Higgs potential. 
In order to parameterise this effect, we introduce the invariant $J_1$ defined in Eq.~(\ref{eq:invariance}). 
In Fig.~\ref{brs} (left), we show the correlations between Im$[\kappa_t]$ and $J_1$. 
We see that, even for small values of $|\text{Im}[\kappa_t]|$,  $J_1$ can be large. 
This can be understood by looking at the Im[$\kappa_t$] plot in Fig.~\ref{fig:kappa}, where the value of Im[$\kappa_t$]
can be close to zero at $\theta_t \neq 0$ due to cancellations. 
As a result, there is a mild correlation between these two observables. 
In contrast, we find a strong correlation between $|\text{Im}[\kappa_t]|$ and the product  BR($H_2 \to VV$) $\times$ BR($H_3 \to VV$), with $V=W^\pm,Z$. As expected,
such a quantity becomes zero in the CPC limit, because either $H_2$ or $H_3$ corresponds to the purely CP-odd state of the C2HDM with CPC, which then does not couple to two gauge bosons (only the CP-even does).  Therefore, this product can measure the CPV effects in the Higgs potential, as already proposed in Ref.~\cite{Keus:2015hva} in the E2HDM with CPV. 
It is seen that the maximal value of this BR product tends to increase with larger $|\text{Im}[\kappa_t]|$ and can be $\sim 2\%$ for $|\text{Im}[\kappa_t]|\sim 5\times 10^{-3}$. 

In short, at the LHC, our scenario can be  probed indirectly by studying the combination of measurements involving $\kappa_V$, Re[$\kappa_t$] and finding two  
additional neutral Higgs bosons decaying into $W^+W^-$ and/or $ZZ$ pairs. In fact, Im[$\kappa_t$] could also potentially  be accessed, e.g., in $t\bar t H_1$ final states \cite{Cheung:2020ugr}.

\section{Summary and Conclusions}
\label{sec:Summary}
In this paper, we have studied a CHM based on
the global symmetry breaking $SO(6)\to SO(4)\times SO(2)$, paying particular attention to CPV generated in the strong sector. 
The 8 pNGBs that emerge from such a dynamics  behave 
as two $SU(2)$ doublet Higgs fields and their 
 properties  are determined by the strong dynamics behind the aforementioned breaking.
We have computed the Higgs potential and top Yukawa coupling generated dynamically in this  
C2HDM by using a two-site moose construction, which effectively describes the strong sector. 
The Higgs potential and Yukawa coupling are expressed in terms of the momentum integration of several form factors which encode the details of the new strong dynamics, chiefly, of the associated CPV. We have found that complex phases in the partial compositeness parameters  can induce CPV in both the Higgs potential and top Yukawa coupling.

We have then discussed typical CPV signatures of such a C2HDM by considering a simple setup, in which there is only one physical complex phase in the strong sector. In the presence
of parameter space constraints obtained from reconstructing the SM VEV as well as the top quark and SM-like Higgs boson masses, we
have first found that $\tan\beta$ should be much larger or lower than 1 to avoid a large $\hat{T}$ parameter while guaranteeing sizable CPV. This is due to the coset structure of $SO(6)/[SO(4)\times SO(2)]$, which cannot forbid $\hat{T}$ contributions at tree level. Next, we
have observed that, once we turn on CPV, the mixing effects amongst the neutral Higgs bosons remain negligible and the masses of extra Higgs bosons become large in comparison to that of the SM-like Higgs boson.  We have also found that the masses of the additional Higgs bosons are nearly degenerate, which in turn  
 implies that they mainly decay into the SM-like Higgs boson (in pairs or in combination with a gauge boson).  Then, 
we have estimated the couplings of the SM-like Higgs boson to the EW gauge bosons and top quark as a function of the CPV phase $\theta_t$. 
Our results are the following: (i) a non-zero CP phase slightly enhances $\kappa_V$; (ii) deviations from the CPC top Yukawa coupling can be of order $10\%$;  (iii) the CPV top Yukawa coupling ($\mbox{Im}[\kappa_t]$) can be $\mathcal{O}(0.01)$. The upper limit of $\mbox{Im}[\kappa_t]$ comes from the requirement  that $v_{\rm{SM}}\simeq 246\,\mbox{GeV}$, $m_t\simeq 173\,\mbox{GeV}$ and $m_{H_1}\simeq 125\,\mbox{GeV}$.
We have further provided analytic expressions for the coupling deviation factors, so-called $\kappa$'s, which clearly justify the above observations. 
In addition, we have discussed the correlation between CPV effects in the top Yukawa coupling and the Higgs potential. In doing so, we have confirmed that there is a mild positive correlation between $\mbox{Im}[\kappa_t]$ and the physical complex phase of the Higgs potential. 
This is one remarkable aspect of the C2HDM with CPV because there should be no such a correlation in the E2HDM with CPV. 
Finally, we have also estimated the correlation between $\mbox{Im}[\kappa_t]$ and the product  BR($H_2 \to VV$) $\times$ BR($H_3 \to VV$), with $V=W^\pm,Z$, as
it is sensitive to the mere presence of CPV in the Higgs potential. 
It was found that the maximal value of this product of BRs depends on the size of $\mbox{Im}[\kappa_t]$: e.g.,  
BR($H_2 \to VV$) $\times$ BR($H_3 \to VV$) can be $\simeq 2\%$ for $|\text{Im}[\kappa_t]|\sim 5\times 10^{-3}$.

Possible outlooks of our initial investigation will include to assess the testability of our C2HDM with CPV at  colliders in great detail, both present and feature ones. Specifically, for what concerns the LHC,  it will be worth to investigate physical processes where interactions between Higgs bosons and the top quark occur like, e.g., $H_{2,3}$ production and decay into $t\bar t$ and/or $t\bar t H_{1,2}$ final states, including exploiting spin and charge asymmetries therein.
It would also be interesting to study CPV in  other coset models, such as $SO(9)/SO(8)$, as, in this case, we expect to obtain more significant CPV effects  in physics observables because this coset structure forbids a large $\hat{T}$ parameter. We leave these developments to future publications.

\section*{Acknowledgments}
We thank Luigi Delle Rose for useful discussions.
S.M. is supported in part through the NExT Institute and the STFC Consolidated Grant No. ST/L000296/1.
This work was supported by JSPS KAKENHI Grant Numbers JP19K14701 and JP21J01070 (R.N.).  
The work of K.Y. was supported in part by the Grant-in-Aid for Early-Career Scientists, No.~19K14714.

\appendix
\section{Fermion Form Factors with CPV}
\label{eq:formfactors}
In this appendix, we give the analytic expressions for the fermion form factors. Here, we consider the $N=2$ case with $N$ denoting the number of spin-1/2 resonances in the strong sector. We include CPV effects in the following calculations. One can find the corresponding expressions in the case of  CPC in Ref.~\cite{DeCurtis:2018zvh}.

We start with Eq.~(\ref{eq:Lfermi}) with $I,J=1,2$: by integrating out $\Psi^{I}$, we obtain
\begin{align}
\tilde{\Pi}^q_0
&=
1-\frac{l_{02}p^2+l_{00}}{p^4-a_0p^2+b_0}
\,,\\
\tilde{\Pi}^q_1
&=
\frac{i}{2}
\biggl(
\frac{-l_{R2}p^2-l_{R0}+l_I}{p^4-(a_R-a_I)p^2+b_R-b_I}
+
\frac{l_{R2}p^2+l_{R0}+l_I}{p^4-(a_R+a_I)p^2+b_R+b_I}
\biggr)
\,,\\
\tilde{\Pi}^q_2
&=
\frac{1}{2}
\biggl(
\frac{l_{R2}p^2+l_{R0}-l_I}{p^4-(a_R-a_I)p^2+b_R-b_I}
+
\frac{l_{R2}p^2+l_{R0}+l_I}{p^4-(a_R+a_I)p^2+b_R+b_I}
\biggr)
-\frac{l_{02}p^2+l_{00}}{p^4-a_0p^2+b_0}
\,,\\
\tilde{\Pi}^t_0
&=
1-\frac{r_{02}p^2+r_{00}}{p^4-a_0p^2+b_0}
\,,\\
\tilde{\Pi}^t_1
&=
\frac{i}{2}
\biggl(
\frac{-r_{R2}p^2-r_{R0}+r_I}{p^4-(a_R-a_I)p^2+b_R-b_I}
+
\frac{r_{R2}p^2+r_{R0}+r_I}{p^4-(a_R+a_I)p^2+b_R+b_I}
\biggr)
\,,\\
\tilde{\Pi}^t_2
&=
\frac{1}{2}
\biggl(
\frac{r_{R2}p^2+r_{R0}-r_I}{p^4-(a_R-a_I)p^2+b_R-b_I}
+
\frac{r_{R2}p^2+r_{R0}+r_I}{p^4-(a_R+a_I)p^2+b_R+b_I}
\biggr)
-\frac{r_{02}p^2+r_{00}}{p^4-a_0p^2+b_0}
\,,\\
\tilde{M_0}
&=
-\frac{m_{02}p^2+m_{00}}{p^4-a_0p^2+b_0}
\,,\\
\tilde{M_1}
&=
\frac{1}{2}
\biggl(
\frac{(m_{12}-im_{22})p^2+m_{10}-im_{20}}{p^4-(a_R-a_I)p^2+b_R-b_I}
+
\frac{(m_{12}+im_{22})p^2+m_{10}+im_{20}}{p^4-(a_R+a_I)p^2+b_R+b_I}
\biggr)
\,,\\
\tilde{M_2}
&=
\frac{1}{2}
\biggl(
\frac{(m_{22}+im_{12})p^2+m_{20}+im_{10}}{p^4-(a_R-a_I)p^2+b_R-b_I}
+
\frac{(m_{22}-im_{12})p^2+m_{20}-im_{10}}{p^4-(a_R+a_I)p^2+b_R+b_I}
\biggr)
-\frac{m_{02}p^2+m_{00}}{p^4-a_0p^2+b_0}
\,,
\end{align}
with $p^2$ being a squared momenta and
\begin{align}
&a_0
=
\mbox{tr}[M^\dag_\Psi M_\Psi]
\,,\label{eq:a0}\\
&a_I
=
2\mbox{Im}\l(\mbox{tr}[Y^\dag_1\bar{Y}_2]\r)
\,,\\
&a_R
=
\mbox{tr}[Y^\dag_1Y_1+\bar{Y}^\dag_2\bar{Y}_2]
\,,\\
&b_0
=
|\mbox{det}M_\Psi|^2
\,,\\
&b_I
=
-2\mbox{Im}\biggl[
(\mbox{det}Y_1-\mbox{det}\bar{Y}_2)\l(
\mbox{tr}[\sigma_2 Y_1\sigma_2 \bar{Y}^T_2]
\r)^*
\biggr]
\,,\\
&b_R
=
|\mbox{det}Y_1-\mbox{det}\bar{Y}_2|^2
+
|\mbox{tr}[\sigma_2 Y_1\sigma_2 \bar{Y}^T_2]|^2
\,,\\
&l_I
=
-2\mbox{Im}\l[
\Delta_L\sigma_2 
(Y^T_1 \bar{Y}^*_2) 
\sigma^T_2\Delta^\dag_L
\r]\,,\\
&r_I
=
-2\mbox{Im}\l[
\Delta_R\sigma_2 
(\bar{Y}^*_2 Y^T_1 ) 
\sigma^T_2\Delta^\dag_R
\r]\,,\\
&l_{02}=
l_{R2}=
\Delta_L\Delta^\dag_L\,,\\
&r_{02}=
r_{R2}=
\Delta_R\Delta^\dag_R
\,,\\
&m_{02}=
\Delta_L M^\dag_\Psi \Delta^\dag_R
\,,\\
&m_{12}=
\Delta_L Y^\dag_1 \Delta^\dag_R
\,,\\
&m_{22}=
\Delta_L \bar{Y}^\dag_2 \Delta^\dag_R
\,,\\
&l_{00}=
\Delta_L\sigma_2\l(M^T_\Psi M^*_\Psi\r)\sigma^T_2\Delta^\dag_L\,,\\
&r_{00}=
\Delta_R\sigma_2\l( M^*_\Psi M^T_\Psi\r)\sigma^T_2\Delta^\dag_R
\,,\\
&l_{R0}=
\Delta_L\sigma_2\l(Y^T_1 Y^*_1 + \bar{Y}^T_2 \bar{Y}^*_2\r)\sigma^T_2\Delta^\dag_L
\,,\\
&r_{R0}=
\Delta_R\sigma_2\l(Y^*_1 Y^T_1 +  \bar{Y}^*_2 \bar{Y}^T_2\r)\sigma^T_2\Delta^\dag_R
\,,\\
&m_{00}=
\Delta_L\sigma_2\l(M^T_\Psi \sigma_2 M^*_\Psi \sigma_2 M^\dag_\Psi\r)\sigma^T_2\Delta^\dag_R
\,,\\
&m_{10}=
\Delta_L\sigma_2\l(
Y^T_1 \sigma_2 Y^*_1 \sigma_2 Y^\dag_1
+
\bar{Y}^T_2 \sigma_2 Y^*_1 \sigma_2 \bar{Y}^\dag_2
+
\bar{Y}^T_2 \sigma_2 \bar{Y}^*_2 \sigma_2 {Y}^\dag_1
-
{Y}^T_1 \sigma_2 \bar{Y}^*_2 \sigma_2 \bar{Y}^\dag_2
\r)\sigma^T_2\Delta^\dag_R
\,,\\
&m_{20}=
\Delta_L\sigma_2\l(
\bar{Y}^T_2 \sigma_2 \bar{Y}^*_2 \sigma_2 \bar{Y}^\dag_2
+
{Y}^T_1 \sigma_2 \bar{Y}^*_2 \sigma_2 {Y}^\dag_1
+
{Y}^T_1 \sigma_2 {Y}^*_1 \sigma_2 \bar{Y}^\dag_2
-
\bar{Y}^T_2 \sigma_2 {Y}^*_1 \sigma_2 {Y}^\dag_1
\r)\sigma^T_2\Delta^\dag_R, 
\label{eq:m20}
\end{align}
with $\bar{Y}_2=M_\Psi-Y_2$.
\section{Higgs Potential Parameters}
\label{app:Higgspotential}
The Higgs potential parameters, $m^2_i$ and $\lambda_i$, are calculated as 
\begin{align}
m^2_i&=
-\frac{i}{f^4}\int \frac{d^4 p}{(2\pi)^4}
\l[\frac{3}{2}(m^G_i)^2-6(m^t_i)^2\r]
\,,~~~~(i=1, ... ,3)\,,\\
\lambda_i&=
-\frac{i}{f^4}\int \frac{d^4 p}{(2\pi)^4}
\l[\frac{3}{2}\lambda^G_i-6\lambda^t_i\r]
\,,~~~~(i=1, ... ,7)\,,
\end{align}
with $(\cdots)^G$ and $(\cdots)^t$ denoting contributions from the gauge bosons and top quark, respectively. Here, we neglect the contributions of all other SM fermions, including the $b$-quark and $\tau$-lepton. Specifically, all such light fermion contributions should be negligibly small because their Yukawa couplings are highly suppressed.

The gauge boson contributions are given in Ref.~\cite{DeCurtis:2018zvh}. We note that the gauge boson contribution preserves  the CP symmetry.  In contrast, the top quark contribution generally breaks the CP symmetry as we see in Eqs.~(\ref{eq:Imm3sqwthetat})--(\ref{eq:Imlam7wthetat}). The corresponding Higgs potential parameters are obtained as
\begin{align}
&\frac{(m^t_1)^2}{f^2}
=
\cos{2\theta_t}\l(\frac{|M_1|^2-|M_2|^2}{2p^2}+\Pi^t_2\r)
+
\sin{2\theta_t}\l(\frac{\mbox{Im}\l[M_2M^*_1\r]}{p^2}-i\Pi^t_1\r)
-
\frac{|M_1|^2+|M_2|^2}{2p^2}-\Pi^q_2+\Pi^t_2
\,,
\label{eq:m1sqwthetat}\\
&\frac{(m^t_2)^2}{f^2}
=
\cos{2\theta_t}\l(\frac{|M_2|^2-|M_1|^2}{2p^2}-\Pi^t_2\r)
+
\sin{2\theta_t}\l(\frac{\mbox{Im}\l[M_2M^*_1\r]}{p^2}-i\Pi^t_1\r)
-
\frac{|M_1|^2+|M_2|^2}{2p^2}-\Pi^q_2+\Pi^t_2
\,,\\
&\frac{\mbox{Re}\l[(m^t_3)^2\r]}{f^2}
=
\cos{2\theta_t}
\frac{\mbox{Re}\l[M_2M^*_1\r]}{p^2}
\,,\\
&{\lambda}^t_1
=
-
\cos{2\theta_t}
\l(
\frac{2|M_1|^2}{3p^2}
-\frac{8|M_2|^2}{3p^2}
+2(\Pi^t_2)^2
+\frac{4}{3}\Pi^t_2
\r)
-
\sin{2\theta_t}
\l(
\frac{10}{3}\frac{\mbox{Im}\l[M_2M^*_1\r]}{p^2}
-
\frac{i}{3}\Pi^t_1
-
2i\Pi^t_1\Pi^t_2
\r)
\nn\\
&
~~~~~~~~~~~~~
+
2i\cos{2\theta_t}\sin{2\theta_t}\Pi^t_1\Pi^t_2
+
\frac{2|M_1|^2}{3p^2}
+\frac{8|M_2|^2}{3p^2}
-(\Pi^t_2)^2
-(\Pi^q_2)^2
-\frac{4}{3}\Pi^t_2
+\frac{4}{3}\Pi^q_2
\,,\\
&{\lambda}^t_2
=
\cos{2\theta_t}
\l(
\frac{2|M_1|^2}{3p^2}
-\frac{8|M_2|^2}{3p^2}
+2(\Pi^t_2)^2
+\frac{4}{3}\Pi^t_2
\r)
-
\sin{2\theta_t}
\l(
\frac{10}{3}\frac{\mbox{Im}\l[M_2M^*_1\r]}{p^2}
-
\frac{i}{3}\Pi^t_1
-
2i\Pi^t_1\Pi^t_2
\r)
\nn\\
&
~~~~~~~~~~~~~
-
2i\cos{2\theta_t}\sin{2\theta_t}\Pi^t_1\Pi^t_2
+
\frac{2|M_1|^2}{3p^2}
+\frac{8|M_2|^2}{3p^2}
-(\Pi^t_2)^2
-(\Pi^q_2)^2
-\frac{4}{3}\Pi^t_2
+\frac{4}{3}\Pi^q_2
\,,\\
&{\lambda}^t_3 
=
-2
\sin{2\theta_t}\l(
\frac{\mbox{Im}\l[M_2M^*_1\r]}{p^2}
-i\Pi^t_1\Pi^t_2
-\frac{i}{2}\Pi^t_1
\r)
+\frac{2|M_1|^2}{p^2}
+(\Pi^q_1)^2
\,,\\
&{\lambda}^t_4 
=
-\frac{2}{3}
\sin{2\theta_t}\l(
\frac{\mbox{Im}\l[M_2M^*_1\r]}{p^2}
+\frac{i}{2}\Pi^t_1
\r)
-\frac{2|M_1|^2}{3p^2}
+\frac{4|M_1|^2}{3p^2}
-(\Pi^q_2)^2
+\frac{2}{3}\Pi^q_2
-\frac{2}{3}\Pi^t_2
\,,\\
&\mbox{Re}\l[{\lambda}^t_5\r]
=
-\frac{2}{3}
\sin{2\theta_t}\l(
\frac{\mbox{Im}\l[M_2M^*_1\r]}{p^2}
+\frac{i}{2}\Pi^t_1
\r)
-\frac{2|M_1|^2}{3p^2}
+\frac{4|M_1|^2}{3p^2}
-(\Pi^q_1)^2
+\frac{2}{3}\Pi^q_2
-\frac{2}{3}\Pi^t_2
\,,\\
&\mbox{Re}\l[{\lambda}^t_6\r]
=
\frac{5\cos{2\theta_t} }{3}{} \frac{\mbox{Re}\left[{M_2} {M^*_1}\right]}{ p^2}
\,,\\
&\mbox{Re}\l[{\lambda}^t_7\r]
=
\frac{5\cos{2\theta_t} }{3 } \frac{\mbox{Re}\left[{M_2} {M^*_1}\right]}{p^2}
\,,
\end{align}
where we ignore cubic terms in the  form factors. 
Notice that $\Pi^q$ and $M_{1,2}$ are defined in Eq.~(\ref{eq:wotildeformfac}) while $\Pi^t$ can be calculated as
\begin{align}
\Pi^{t}_{1,2}
=
\frac{\tilde{\Pi}^{t}_{1,2}}{\tilde{\Pi}^{t}_0-\tilde{\Pi}^{t}_2+i\sin{2\theta_t}\tilde{\Pi}^t_1}\,.
\end{align}

It should also be noted that the momentum integrations for $\Pi_{1,2}^{q}$ and $\Pi_{1,2}^{t}$ have a UV divergence because they behave as $\sim p^{-4}$ in the large $p^2$ regime. We find that, in order to ensure UV finiteness, we need to impose the following conditions:
\begin{align}
&a_0I_{02}+l_{00}=a_Rl_{02}+l_{R0}\,~~~(\mbox{for}~\Pi^q_2)\,,\\
&a_0r_{02}+r_{00}=a_Rr_{02}+r_{R0}\,~~~(\mbox{for}~\Pi^t_2)\,,\\
&a_Il_{02}=-l_{I}\,~~~(\mbox{for}~\Pi^q_1)\,,\\
&a_Ir_{02}=-r_{I}\,~~~(\mbox{for}~\Pi^t_1)\,,
\end{align}
where $a$, $l$ and $r$ are defined in Eqs.~(\ref{eq:a0})--(\ref{eq:m20}).
For the case where the number of spin-1/2 resonances is two, these conditions are equivalent to, respectively, 
\begin{align}
&(\Delta_L\Delta^\dag_L)
\mbox{tr}[Y^T_1 Y^*_1+\bar{Y}^T_2 \bar{Y}^*_2-M^T_\Psi M^*_\Psi]
=
\Delta_L\sigma_2\l(M^T_\Psi M^*_\Psi-Y^T_1 Y^*_1 - \bar{Y}^T_2 \bar{Y}^*_2\r)\sigma^T_2\Delta^\dag_L
\,~~~(\mbox{for}~\Pi^q_2)\,,\\
&
(\Delta_R\Delta^\dag_R)
\mbox{tr}[ Y^*_1 Y^T_1+\bar{Y}^*_2 \bar{Y}^T_2 -M^*_\Psi M^T_\Psi]
=
\Delta_R\sigma_2\l(M^*_\Psi M^T_\Psi- Y^*_1 Y^T_1-\bar{Y}^*_2 \bar{Y}^T_2 \r)\sigma^T_2\Delta^\dag_R
\,~~~(\mbox{for}~\Pi^t_2)\,,\\
&
(\Delta_L\Delta^\dag_L)\mbox{Im}\l(\mbox{tr}[Y^\dag_1 \bar{Y}_2]\r)
=\mbox{Im}\l(\Delta_L\sigma_2\l(Y^T_1 \bar{Y}^*_2\r)\sigma^T_2\Delta^\dag_L\r)
\,~~~(\mbox{for}~\Pi^q_1)\,,\\
&
(\Delta_R\Delta^\dag_R)\mbox{Im}\l(\mbox{tr}[Y^\dag_1 \bar{Y}_2]\r)
=\mbox{Im}\l(\Delta_R\sigma_2\l(\bar{Y}^*_2Y^T_1 \r)\sigma^T_2\Delta^\dag_R\r)
\,~~~(\mbox{for}~\Pi^t_1)\,.
\end{align}
We can easily check that 
the UV finiteness conditions are automatically satisfied with the choice of Eq.~(\ref{eq:LRsym}).

In the special case where the CPV source is given only by the $\theta_t$ parameter,  
the top quark contributions to the Higgs potential parameters are obtained as 
\begin{align}
\begin{split}
&(m^t_1)^2
=
\frac{6i}{f^2}\int\frac{d^4 p}{(2\pi)^4}
\biggl[
\cos{2\theta_t}\l(\frac{|M_1|^2-|M_2|^2}{2p^2}+\Pi^t_2\r)
-
\frac{|M_1|^2+|M_2|^2}{2p^2}-\Pi^q_2+\Pi^t_2
\biggr]
\,,\\
&(m^t_2)^2
=
\frac{6i}{f^2}\int\frac{d^4 p}{(2\pi)^4}
\biggl[
\cos{2\theta_t}\l(\frac{|M_2|^2-|M_1|^2}{2p^2}-\Pi^t_2\r)
-
\frac{|M_1|^2+|M_2|^2}{2p^2}-\Pi^q_2+\Pi^t_2
\biggr]
\,,\\
&{\mbox{Re}\l[(m^t_3)^2\r]}
=
\frac{6i}{f^2}\int\frac{d^4 p}{(2\pi)^4}
\biggl[
\cos{2\theta_t}
\frac{\mbox{Re}\l[M_2M^*_1\r]}{p^2}
\biggr]
\,,\\
&{\mbox{Im}\l[(m^t_3)^2\r]}
=
-\frac{6i}{f^2}\int\frac{d^4 p}{(2\pi)^4}
\biggl[
\frac{\sin2\theta_t}{2}\l(\frac{|M_1|^2+|M_2|^2}{p^2}\r)
\biggr]
\,,
\\
&{\lambda}^t_1
=
\frac{6i}{f^4}\int\frac{d^4 p}{(2\pi)^4}
\biggl[
-
\cos{2\theta_t}
\l(
\frac{2|M_1|^2}{3p^2}
-\frac{8|M_2|^2}{3p^2}
+2(\Pi^t_2)^2
+\frac{4}{3}\Pi^t_2
\r)
\\
&
~~~~~~~~~~~~~~~~~~~~~~~
+
\frac{2|M_1|^2}{3p^2}
+\frac{8|M_2|^2}{3p^2}
-(\Pi^t_2)^2
-(\Pi^q_2)^2
-\frac{4}{3}\Pi^t_2
+\frac{4}{3}\Pi^q_2
\biggr]
\,,\\
&{\lambda}^t_2
=
\frac{6i}{f^4}\int\frac{d^4 p}{(2\pi)^4}
\biggl[
\cos{2\theta_t}
\l(
\frac{2|M_1|^2}{3p^2}
-\frac{8|M_2|^2}{3p^2}
+2(\Pi^t_2)^2
+\frac{4}{3}\Pi^t_2
\r)
\\
&
~~~~~~~~~~~~~~~~~~~~~~~
+
\frac{2|M_1|^2}{3p^2}
+\frac{8|M_2|^2}{3p^2}
-(\Pi^t_2)^2
-(\Pi^q_2)^2
-\frac{4}{3}\Pi^t_2
+\frac{4}{3}\Pi^q_2
\biggr]
\,,\\
&{\lambda}^t_3 
=
\frac{6i}{f^4}\int\frac{d^4 p}{(2\pi)^4}
\biggl[
\frac{2|M_1|^2}{p^2}
+(\Pi^q_1)^2
\biggr]
\,,\\
&{\lambda}^t_4 
=
\frac{6i}{f^4}\int\frac{d^4 p}{(2\pi)^4}
\biggl[
-\frac{2|M_1|^2}{3p^2}
+\frac{4|M_1|^2}{3p^2}
-(\Pi^q_2)^2
+\frac{2}{3}\Pi^q_2
-\frac{2}{3}\Pi^t_2
\biggr]
\,,\\
&\mbox{Re}\l[{\lambda}^t_5\r]
=
\frac{6i}{f^4}\int\frac{d^4 p}{(2\pi)^4}
\biggl[
-\frac{2|M_1|^2}{3p^2}
+\frac{4|M_1|^2}{3p^2}
+\frac{2}{3}\Pi^q_2
-\frac{2}{3}\Pi^t_2
\biggr]
\,,\\
&\mbox{Re}\l[{\lambda}^t_6\r]
=
\mbox{Re}\l[{\lambda}^t_7\r]
=
\frac{6i}{f^4}\int\frac{d^4 p}{(2\pi)^4}
\biggl[
\frac{5\cos{2\theta_t} }{3}{} \frac{\mbox{Re}\left[{M_2} {M^*_1}\right]}{ p^2}
\biggr]
\,,\\
&\mbox{Im}\l[{\lambda}^t_6\r]
=
\mbox{Im}\l[{\lambda}^t_7\r]
=
-
\frac{6i}{f^4}\int\frac{d^4 p}{(2\pi)^4}
\biggl[
\frac{2 \sin2\theta_t}{3}\l(\frac{{|M_1|^2+|M_2|^2}}{p^2} \r)
\biggr]
\,,
\end{split} \label{eq:ff-special}
\end{align}
where 
\begin{align}
&\Pi^{q}_{2}
=
\frac{\tilde{\Pi}^{q}_{2}}{\tilde{\Pi}^{q}_0}
\,,
~~
M_{1,2}
=
M^*_{1,2}
=
\frac{\tilde{M}_{1,2}}{\sqrt{\tilde{\Pi}^{q}_0(\tilde{\Pi}^{t}_0-\tilde{\Pi}^{t}_2)}}\,.
\end{align}

\bibliography{ref_CPVC2HDM} 
\bibliographystyle{JHEP}
\end{document}